\documentclass[12pt]{article}
\usepackage{amscd,amssymb,amsmath,latexsym,enumerate}
\usepackage[mathscr]{euscript}
\usepackage{epsfig}
\usepackage{fancybox}
\usepackage{verbatim}
\usepackage{multirow}
\usepackage{stmaryrd}
\usepackage{bm}

\usepackage[hidelinks]{hyperref}
\hypersetup{ 
       colorlinks=true,
       linkcolor=black, 
       citecolor=black,
       filecolor=black,
       menucolor=black,
       urlcolor=black,
       breaklinks=true
}

\usepackage{color}

\title{
Topological indices for periodic gapped Hamiltonians \\ and fuzzy tori
}

\author{Nora Doll$^1$, Terry Loring$^2$, Hermann Schulz-Baldes$^3$
\\
\\
{\small $^1$Institut f\"ur Mathematik, Martin-Luther-Universit\"at Halle-Wittenberg, Germany}
\\
{\small $^2$Department of Mathematics and Statistics, University of New Mexico, Albuquerque, USA}
\\
{\small $^3$Department Mathematik, Friedrich-Alexander-Universit\"at Erlangen-N\"urnberg, Germany}
\\
{\small $^3$corresponding author, schuba@mi.uni-erlangen.de}
}

\date{ }

\newtheorem{theo}{Theorem}
\newtheorem{defini}[theo]{Definition}
\newtheorem{proposi}[theo]{Proposition}
\newtheorem{lemma}[theo]{Lemma}
\newtheorem{coro}[theo]{Corollary}
\newtheorem{remark}[theo]{Remark}
\newtheorem{example}[theo]{Example}

\newcommand{\MM}{{\mathbb M}}
\newcommand{\BM}{{\mathbb B}}
\newcommand{\CM}{{\mathbb C}}
\newcommand{\NM}{{\mathbb N}}
\newcommand{\RM}{{\mathbb R}}
\newcommand{\SM}{{\mathbb S}}
\newcommand{\TM}{{\mathbb T}}
\newcommand{\ZM}{{\mathbb Z}}

\newcommand{\UM}{{\mathbb U}}

\newcommand{\Aa}{{\cal A}}

\newcommand{\Ff}{{\cal F}}

\newcommand{\Oo}{{\cal O}}

\newcommand{\Rr}{{\cal R}}

\newcommand{\Mm}{{\cal M}}

\newcommand{\Kk}{{\cal K}}
\newcommand{\Hh}{{\cal H}}

\newcommand{\one}{{\bf 1}}

\newcommand{\per}{{\mbox{\rm\tiny per}}}

\newcommand{\Tr}{\mbox{\rm Tr}}

\newcommand{\Ch}{{\rm Ch}} 
\newcommand{\Ind}{{\rm Ind}} 
\newcommand{\Ker}{{\rm Ker}}

\newcommand{\sgn}{{\rm sgn}} 
\newcommand{\Sig}{{\rm Sig}}

\newcommand{\Pf}{{\rm Pf}}

\newcommand{\gah}{\hat{\gamma}} 
\newcommand{\gao}{\gamma}

\newcommand{\spec}{{\rm spec}}

\newcommand{\PL}{L^{\mbox{\rm\tiny per}}}
\newcommand{\SPL}{L^{\mbox{\rm\tiny skew,per}}}
\newcommand{\skewu}{\mbox{\rm\tiny skew}}

\begin{document}

\maketitle

\begin{abstract} 
A variety of local index formulas is constructed for gapped quantum Hamiltonians with periodic boundary conditions. All dimensions of physical space as well as many symmetry constraints are covered, notably one-dimensional systems in Class DIII as well as two- and three-dimensional systems in Class AII. The constructions are based on several periodic variations of the spectral localizer and are rooted in the existence of underlying fuzzy tori. For these latter, a general invariant theory is developed. 

\vspace{.1cm}

\noindent MSC 2010: 81R60, 37B30, 46L80, 81V70, 47S40  
\end{abstract}



\vspace{-.7cm}

\section{Overview}
\label{sec-PL}

In several works \cite{LS1,LS2,DSW} it was shown that the index pairing between $K$-theory and $K$-homology elements can be computed as the half-signature of a suitably constructed finite-dimensional matrix, called the spectral localizer. The main motivation for these works is the application to topological insulators for which the bulk topological invariants (Chern numbers and winding numbers) then become readily accessible in numerical computations. As will be described below, the technique is based on  the principle of placing the physical system in a linearly growing Dirac trap and hence the spectral localizer is an intrinsically non-periodic object. On the other hand, it is well-known that periodic approximations often provide stable algorithms for bulk quantities in solid state systems (for invariants, this is described in \cite{Pro}). This work constructs new periodic versions of the spectral localizer, for sake of conciseness referred to as {\it periodic spectral localizers}, which also allow to compute the topological invariants numerically, possibly in a more stable manner than with the non-periodic spectral localizer used in other works. Apart from this practical aspect, a further more theoretical insight is that the periodic spectral localizers can be understood as the $K$-theory representatives of associated fuzzy tori. Furthermore, the periodic spectral localizers may inspire extensions to interacting systems (with periodic boundary conditions) with a computable gapped ground state.

\vspace{.2cm}

Let us directly describe the periodic spectral localizer for a bounded tight-binding Hamiltonian $H=H^*$ on the Hilbert space $\Hh=\ell^2(\ZM^d,\CM^L)$ over an even dimensional lattice $\ZM^d$ with $L$ local degrees of freedom. This Hamiltonian is supposed to satisfying the following hypothesis:

\vspace{.2cm}

\noindent $\bullet$ $H$ has finite range $R$, namely the $L\times L$ matrices $\langle x|H|y\rangle$ vanish for $|x-y|>R$; 

\vspace{.1cm}

\noindent $\bullet$ $H$ is periodic, namely  $\langle x+p_j e_j|H|y+p_je_j\rangle=\langle x|H|y\rangle$ for some periods $p_j\in\NM$ in the direction 

of the unit vectors $e_j$, $j=1,\ldots,d$; 

\vspace{.1cm}

\noindent $\bullet$ $0$ lies in a spectral gap of $H$.

\vspace{.2cm}

\noindent In physical terminology, the last condition means that $H$ describes an insulator (here the Fermi level is shifted to $\mu=0$). For any such insulator it is well-known ({\it e.g.} \cite{PSB}) that the Fermi projection $P=\chi(H<0)$ has an associated (even strong) Chern number $\Ch_d(P)\in\ZM$. This paper provides yet another way to compute this topological invariant. Moreover, the formula that is proven to work in the asymptotic regime of large volumes also allows to associate numerical topological invariants to rather small systems with periodic boundary conditions. This enables efficient algorithms that can analyze the effect of defects or disorder on the bulk invariants without the confounding influence of edge effects.

\begin{defini}
\label{eq-defHPer}
Let $p_1,\ldots,p_d$ be the periods a short range Hamiltonian $H=H^*$ on $\Hh$. Suppose that $\rho\in\NM$ is such that $2\rho$ is an integer multiple of each of the $p_j$, and denote $\Hh_\rho=\ell^2((\ZM/(2\rho\ZM))^d,\CM^L)$. The finite-volume restriction $H^\per_\rho$ of $H$ with periodic boundary conditions is an operator on $\Hh_\rho$ defined by
$$
\langle x|H^\per_\rho|y\rangle
\;=\;
\sum_{a\in\ZM^d} \langle x|H|y+2\rho a\rangle
\;,
\qquad
x,y\in(\ZM/2\rho \ZM)^d\cong\{-\rho+1,\ldots,\rho\}^d 
\;.
$$
\end{defini}

Note that the sum over $a\in\ZM^d$ is finite due to the finite range assumption.

\begin{defini}
Let $d$ be even and $H$, $H^\per_\rho$ and $\Hh_\rho$ as above. Further let $\eta>0$. The  (even) periodic spectral localizer is a finite-dimensional matrix on $(\Hh_\rho\oplus\Hh_\rho)\otimes \CM^{d'}$ defined by
\begin{equation}
\label{eq-PLintro}
\PL_{\eta,\rho}
\,=\,
\begin{pmatrix}
\sum_{j=1}^{d} \big(\one-\cos(\tfrac{\pi}{\rho}\, X_j)\big)
&
\sum_{j=1}^{d} \sin(\tfrac{\pi}{\rho}\, X_j)\,\gah_j^*
\\
\sum_{j=1}^{d} \sin(\tfrac{\pi}{\rho}\, X_j)\,\gah_j
&
-\sum_{j=1}^{d} \big(\one-\cos(\tfrac{\pi}{\rho}\, X_j)\big)
\end{pmatrix}
+
\frac{1}{\eta}
\begin{pmatrix}
-H^\per_\rho & 0 \\ 0 & H^\per_\rho
\end{pmatrix}
\;,
\end{equation}
where $X_j$ are the components $j=1,\ldots,d$ of the position operators on the lattice, furthermore $\gah_1,\ldots,\gah_{d-1}$ is a selfadjoint irreducible representation on $\CM^{d'}$ of the Clifford algebra with $d-1$ generators, namely $\gah_i\gah_j+\gah_j\gah_i=2\,\delta_{i,j}$ for $i,j=1,\ldots,d-1$, and $\gah_d=\imath \one$ with $\imath=\sqrt{-1}$. 
\end{defini}

In Remark~\ref{rem-EtaDiscuss} below it is argued that $\eta$ should roughly be chosen as $\|H\|$, and hence not be considered a free parameter in applications. Note that the first summand in \eqref{eq-PLintro} is the restriction of a diagonal operator onto $(\Hh_\rho\oplus\Hh_\rho)\otimes \CM^{d'}$ and it is periodic in all $d$ directions of the discrete torus $(\ZM / (2\rho\ZM))^d\cong \ZM^d\cap [-\rho+1,\rho]^d$. The following result now states that the periodic spectral localizer has, under suitable conditions, a well-defined signature that is equal to the Chern number. On first sight this may look like a minor modification of earlier results \cite{LS2,DSW}. However, the main player $\PL_{\eta,\rho}$ is different here. It is periodic by construction, potentially useful at very small volumes $\rho$ and allows to establish deep connections to invariants of associated fuzzy tori. All of these points will be discussed in detail in the remainder of this introduction.

\begin{theo}
\label{theo-Intro}
Let $d$ be even and $H=H^*$ be a finite-range periodic operator on $\Hh=\ell^2(\ZM^d,\CM^L)$. Also let $\rho\in \NM$ be such that $2\rho$ is an integer multiple of the periods and $\rho\geq 2 R$. Suppose
\begin{align}
\label{eq-IntroCond}
\rho
\;\geq\;
\frac{C\,d^4\,M\,\|H\|^3\,\eta^2}{g^6}
\;,
\end{align}
where $g=\|H^{-1}\|^{-1}$, $M=\max_{j=1,\ldots,d}\|[X_j,H]\|$, and finally $C=15\cdot10^6$. Moreover,  $\eta\geq \frac{g}{4}$ is such that
\begin{align}
\label{eq-IntroCond2}
\Big(1-\frac{g}{\|H\|}\Big)^2+4\Big(1-\frac{\eta}{\|H\|}\Big)\,\leq\,\frac{g^2}{4\,d\,\eta\,\|H\|}
\;.
\end{align}
Then  the periodic spectral localizer is gapped with lower bound
\begin{equation}
\label{eq-PLgap}
(\PL_{\eta,\rho})^2\;\geq\;
\frac{g^2}{600\,d\,\eta^2}\,\one
\;,
\end{equation}
and the strong invariant given by the $d$-th Chern number of $P=\chi (H< 0)$ is equal to the half-signature of the periodic spectral localizer, namely
\begin{equation}
\label{eq-EvenIntro}
\Ch_d(P)
\;=\;
\frac{1}{2}\,\Sig(\PL_{\eta,\rho})
\;.
\end{equation}
\end{theo}

\begin{remark}
\label{rem-SpecLocComp}
{\rm
Let us start out by comparing the periodic spectral localizer $\PL_{\eta,\rho}$ with the spectral localizer $L_{\kappa,\rho}$ used in prior works \cite{LS1,LS2,DSW}, and also explain the connection between the two of them. The latter matrix $L_{\kappa,\rho}$ is defined on the same finite-dimensional Hilbert space $(\Hh_\rho\oplus\Hh_\rho)\otimes \CM^{d'}$ by
\begin{equation}
\label{eq-SLintro}
L_{\kappa,\rho}
\;=\;
\kappa 
\begin{pmatrix}
0 & \sum_{j=1}^{d}X_j\gah_j^*\\ \sum_{j=1}^{d}X_j\gah_j & 0
\end{pmatrix}
\,+\,
\begin{pmatrix}
-H_\rho & 0 \\ 0 & H_\rho
\end{pmatrix}
\;.
\end{equation}
Here $H_\rho$ is the restriction of $H$ to $\Hh_\rho$, also called either the compression of $H$ or $H$ with Dirichlet boundary conditions. The first matrix, without the factor $\kappa$, is called the Dirac operator $D$ and is off-diagonal as the dimension $d$ is even. The spectral localizer $L_{\kappa,\rho}$ is clearly not periodic in the above sense because the position operators take large positive and negative values at the boundaries of the discrete cube $\ZM^d\cap [-\rho+1,\rho]^d$.  The main result of \cite{LS2} (see also \cite{DSW}) is that the equality \eqref{eq-EvenIntro} holds with  $\PL_{\eta,\rho}$ on the r.h.s. replaced by $L_{\kappa,\rho}$, provided conditions on $\kappa$ and $\rho$ hold that are quantitatively weaker than \eqref{eq-IntroCond}. In the latter regime, the spectral asymmetry of both operators $\PL_{\eta,\rho}$ and $L_{\kappa,\rho}$ is acquired near the center of the finite volume where both operators are roughly the same which gives an intuitive understanding why Theorem~\ref{theo-Intro} should hold (based on the earlier results \cite{LS1,LS2}). Indeed, the proof of Theorem~\ref{theo-Intro} consists of constructing a homotopy from $L_{\kappa,\rho}$ to $\PL_{\eta,\rho}$ inside the finite-dimensional invertible selfadjoint matrices. The essential step is a deformation of the first summand in \eqref{eq-PLintro}, which is explicitly given in \eqref{eq-PathTwist} below. It further results from the strategy of proof in Section~\ref{sec-EvenPL} that the Hamiltonian can be tapered down, see \eqref{eq-PLlocalized}. This means that in the regime of \eqref{eq-IntroCond} the contribution to the signature results merely from the central part of the finite volume. Hence the half-signature in \eqref{eq-EvenIntro} is a {\it local topological marker} in this regime, just as the half-signature of the spectral localizer of \cite{LS1,LS2}.  However, one can use the r.h.s. of \eqref{eq-EvenIntro} also for much smaller $\rho$ for which periodic boundary conditions {\it are} relevant so that the signature invariant is a {\it global or bulk topological invariant}. All of this is numerically confirmed in Remark~\ref{rem-Numerics} on the example of a one-dimensional topological system, but there is definitely a need for further investigations.
\hfill $\diamond$}
\end{remark}

\vspace{-.7cm}

\begin{remark}
\label{rem-EtaDiscuss}
{\rm
Besides being the volume, the parameter $\rho$ sets the length scale of the position operator close to the origin because $\sin(\frac{\pi}{\rho} x_j)\sim \frac{\pi}{\rho} x_j$ and $1-\cos(\frac{\pi}{\rho} x_j)\sim \frac{1}{2}(\frac{\pi}{\rho} x_j)^2$. Comparing with the (non-periodic) spectral localizer (spelled out in \eqref{eq-SLintro}), $\frac{1}{\rho}$ hence plays the same role as the parameter $\kappa$ in prior works  \cite{LS1,LS2,DSW}. Having this in mind, the condition in \eqref{eq-IntroCond} is a more stringent version of the main hypothesis in these works. Note that, given a gapped Hamiltonian $H$, it can always be guaranteed by choosing $\rho$ sufficiently large. The second bound \eqref{eq-IntroCond2} is a new supplementary condition. For a flat band Hamiltonian which by definition satisfies $g=\|H\|$, the condition becomes $\eta\geq \frac{g}{2}(1+\sqrt{1-\frac{1}{4d}})$. In particular, $\eta=1$ is allowed for a flat band Hamiltonian with $g=\|H\|=1$. As will be explained in the second part of this introduction and Section~\ref{sec-Fuzzy}, this is reminiscent of the fact that the flat band Hamiltonian together with suitable functions of the position operators forms a {\it graded fuzzy torus}.  On the other hand, if one chooses $\eta=\|H\|$ (without imposing the flat band condition), then the bound \eqref{eq-IntroCond2} becomes $\frac{\|H\|}{g}-1\leq \frac{1}{2\sqrt{d}}$ which means that $H$ has to be relatively close to a flat band Hamiltonian in the sense that $\|H\|$ is not allowed to be much bigger than $g=\|H^{-1}\|^{-1}$. Furthermore, \eqref{eq-IntroCond2} is always satisfied if $\eta\geq \frac{5}{4}\|H\|$ (simply because then the l.h.s. becomes negative due to $g\leq\|H\|$). Note, however, that for large $\eta$, 
\eqref{eq-IntroCond} enforces $\rho$ to be larger, and furthermore the gap of $\PL_{\eta,\rho}$ closes, see the bound \eqref{eq-PLgap}, and then its signature may not be numerically stable any more. Hence from a numerical perspective, it may be best to choose $\eta$ of the order of $\|H\|$ and actually somewhat smaller than $\|H\|$ so that the periodic spectral localizer is associated to a fuzzy torus of small width (in the sense of Definition~\ref{def-Graded} below). In conclusion, the discussion shows that one should  chose $\eta\approx\|H\|$ and in the sequel not consider it as a free parameter. 
\hfill $\diamond$}
\end{remark}

\vspace{-.7cm}

\begin{remark}
\label{rem-ConditionDiscuss}
{\rm
The bounds \eqref{eq-IntroCond} and \eqref{eq-IntroCond2} have an intrinsic scale invariance. Actually, replacing $H$, $M$, $g$ and $\eta$ by $\lambda H$, $\lambda M$, $\lambda g$ and $\lambda \eta$ respectively where $\lambda>0$ is a scaling parameter, leaves the conditions invariant. All four quantities are expressed in energy units, while $\rho$ is a space unit. From a quantitative aspect, we believe that hypothesis \eqref{eq-IntroCond2} is relatively close to optimal, while the condition  \eqref{eq-IntroCond} is certainly off by several orders of magnitude and even the dependence on $g$ is likely much worse than needed. Let us stress that once $\eta$ is chosen as in Remark~\ref{rem-EtaDiscuss}, there is no further parameter other than  $\rho$. One can then analyze numerically the behavior for small $\rho$ and safely use the half-signature as local topological marker, as long as it is stable.
\hfill $\diamond$}
\end{remark}


\begin{remark}
{\rm
As pointed out in Remark~\ref{rem-EtaDiscuss}, the condition \eqref{eq-IntroCond2} is easiest to satisfy if $H$ is already somewhat close to a flat band Hamiltonian. This can be attained by replacing a given initial gapped finite-range periodic Hamiltonian $H'$ by a suitable polynomial $H=q(H')$ which is then also periodic and of finite range, even though the range is increased by a factor given by the degree of the polynomial $q$. The polynomial $q$ should be chosen odd with $q(x)>0$ for $x>0$ so that $H$ also has a spectral gap at $0$, and, moreover, to have a degree as small as possible. Based on the spectral information of $H'$, it is straightforward to construct a suitable polynomial.
\hfill $\diamond$}
\end{remark}

\begin{figure}
\centering
\includegraphics[width=5.1cm,height=4.5cm]{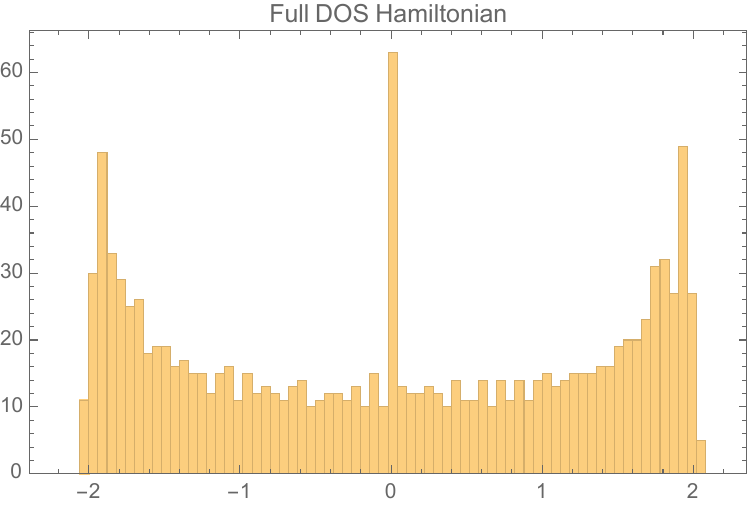}
\hspace{.3cm}
\includegraphics[width=5.1cm,height=4.5cm]{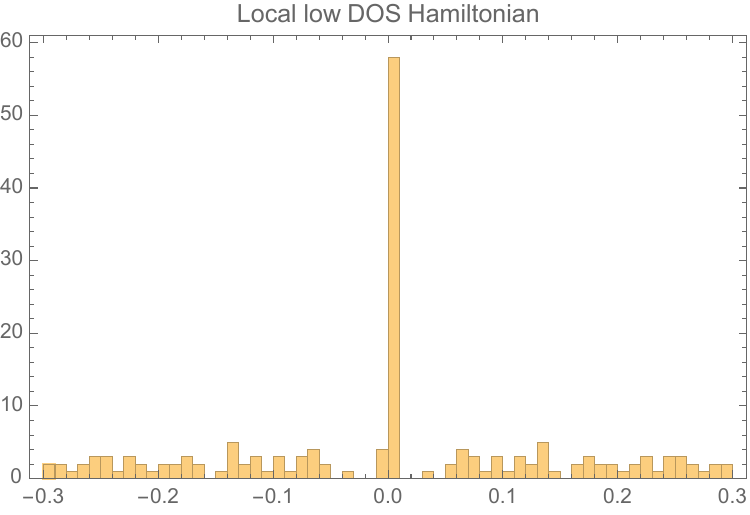}
\hspace{.3cm}
\includegraphics[width=5.1cm,height=4.5cm]{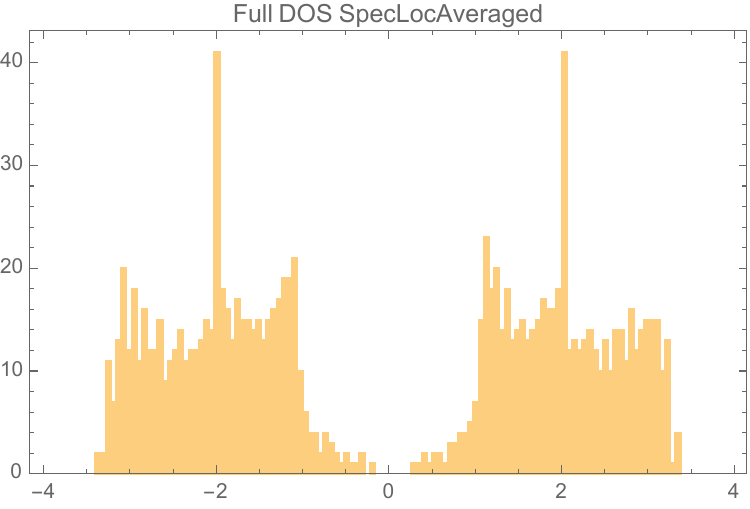}
\caption{\sl 
Plots of the full eigenvalue distribution for one realization of the Hamiltonian $\widetilde{H}^\per_\rho$ described in {\rm Remark~\ref{rem-Numerics}}, then its central part and in the third plot the eigenvalue distribution of ${L}^\per_{1,\rho}$.  The random variables are uniformly distributed, $\lambda=0.5$, $m=0.9\,\imath$, $\rho=300$ and $s=0.05$. Hence the Hilbert space is of dimension $1200$ states and due to $s=0.05$ the kernel of $\widetilde{H}^\per_\rho$ has about $60$ states, and ${L}^\per_{1,\rho}$ has about $30$ supplementary eigenvalues close to $-2$ and $2$ each.  The half-signature of ${L}^\per_{1,\rho}$ is still $-1$.
}
\label{fig-SSH}
\end{figure}

\begin{remark}
\label{rem-Numerics}
{\rm
As already discussed in Remarks~\ref{rem-SpecLocComp} and \ref{rem-ConditionDiscuss}, we expect the signature of $\PL_{\eta,\rho}$ to be stable for much smaller system sizes $\rho$. Let us support this belief by some numerics (see Fig.~\ref{fig-SSH}) in the numerically most simple situation of a chiral model in dimension $d=1$, namely the so-called SSH model (a more detailed description of this much studied object can be found in \cite{PSB}). The chiral Hamiltonian is then an off-diagonal $2\times 2$ matrix with off-diagonal entry $A$ given by an invertible tight-binding operator on $\ell^2(\ZM)$ which in Dirac Bra-Ket notation is given by $A|x\rangle=(m+m_x)||x\rangle+(1+t_x)|x+1\rangle$, where $m\in\CM$, $(m_x)_{x\in\ZM}$ and $(t_x)_{x\in\ZM}$ are independent and identically distributed random variables in $[-\lambda,\lambda]$ with $\lambda<1$. For $\lambda$ and $m$ sufficiently small, the  operator $A$ has a non-commutative winding number equal to $-1$ in the present situation. According to Theorem~\ref{theo-PLOdd} (the odd-dimensional equivalent to Theorem~\ref{theo-Intro}), it can be computed using the following odd periodic spectral localizer discussed in Section~\ref{sec-OddPL}:
\begin{equation}
\label{eq-PLApprox}
{L}^\per_{\eta,\rho}
\;=\;
\begin{pmatrix}
\sin(\frac{\pi}{\rho}X) & \one-\cos(\frac{\pi}{\rho}X  )
\\
\one-\cos(\frac{\pi}{\rho}X) & 
-\sin(\frac{\pi}{\rho}X)
\end{pmatrix}
\,-\,
\frac{1}{\eta}
\begin{pmatrix}
0 & A^\per_\rho \\ (A^\per_\rho)^* & 0
\end{pmatrix}
\;.
\end{equation}
Here $X=X_1$ is the position operator and the right summand is the periodized SSH Hamiltonian $H^\per_\rho$ in finite volume. Numerics readily show that its half-signature is $-1$ for a large span of $\rho$ and $\lambda$, simply confirming  Theorem~\ref{theo-PLOdd}. If periodic boundary conditions are used, this reliably works for $\rho$ as small as $4$. Also interesting is that drastic modifications of the Hamiltonian do not alter the signature index if $\rho$ is sufficiently large: instead of $H^\per_\rho$, one can use $\widetilde{H}^\per_\rho$ obtained from $H^\per_\rho$ by setting all matrix elements $\langle x|H^\per_\rho|y\rangle=0$ for either $|x|>{(1-s)}\rho$ or $|y|>{(1-s)}\rho$ where $s\in[0,1)$. Any $s>0$ will eliminate the periodic boundary conditions and, moreover, leads to a large kernel of $\widetilde{H}^\per_\rho$, roughly of dimension ${4\rho s}$. Nevertheless, this kernel results from regions where the first summand in \eqref{eq-PLApprox} has large off-diagonal parts and hence does not lead to a kernel of ${L}^\per_{\eta,\rho}$. In the central region $[-{(1-s)}\rho,{(1-s)}\rho]\cap\ZM$ where ${L}^\per_{\eta,\rho}$ extracts the topology by means of its spectral asymmetry, the modified Hamiltonian $\widetilde{H}^\per_\rho$ coincides with the ${H}^\per_\rho$. Indeed, numerics clearly show that the half-signature is still $-1$ as long as ${(1-s)\rho\approx 30}$. This clearly shows that the periodic spectral localizer in the regime of large $\rho$ reads out the topology locally close to the origin (where $\sin$ is linear and $1-\cos$ vanishes). 
\hfill $\diamond$}
\end{remark}

\vspace{-.4cm}

\begin{remark}
{\rm
In Section~\ref{sec-Z2PL}, it is shown how to deal with the strong $\ZM_2$-invariants for Hamiltonians lying in the suitable Cartan-Altland-Zirnbauer symmetry classes. In principle, one can also access weak invariants by the techniques of the present work. Indeed, the experienced reader will easily  locate the relevant formulas for fuzzy tori in Section~\ref{sec-Fuzzy}, but this is here not explained in detail for solid state applications. 
\hfill $\diamond$}
\end{remark}


\begin{remark}
{\rm
Theorem~\ref{theo-Intro} is stated for periodic Hamiltonians. However, for space homogeneous random operators (in the sense of Bellissard \cite{Bel}, see also \cite{PSB}), there is a natural construction of periodic approximants \cite{Pro}. For sufficiently large sizes of these approximants, the Chern numbers coincide with those at infinite volume and can be computed using periodic boundary conditions \cite{Pro}. In order to avoid introducing the notational machinery, these results are not spelled out in detail.
\hfill $\diamond$}
\end{remark}


\begin{remark}
{\rm
From a $KK$-theoretic perspective (explained in some detail elsewhere), both $\PL_{\eta,\rho}$ and $L_{\kappa,\rho}$ are representatives of a Kasparov product of two even $KK$-cycles  $[P]\in KK^0(\CM,\Aa)$ and $[D]\in KK^0(\Aa,\CM)$ for a suitable algebra $\Aa$. The two gradings are apparent in \eqref{eq-SLintro}, but once the  Kasparov product is computed, one can deform it without respecting the grading of the separate Fredholm modules $[D]$ and $[P]$. This is reflected by the lack of grading of the first summand in expression \eqref{eq-PLintro} of the periodic spectral localizer.
\hfill $\diamond$}
\end{remark}

Let us conclude this introduction with a brief discussion of the notion of a fuzzy $d$-torus in connection with the periodic spectral localizer. This is discussed in detail in Section~\ref{sec-Fuzzy} which we believe to be of considerable independent interest, possibly  serving as a guideline to the construction of numerically computable local index formulas for other fuzzy versions of classical geometric objects. Abstract index formulas (not suitable for numerical implementation) have been known for a long time \cite{EL,ELo,ELP}, and invariants for the special case of fuzzy spheres were already studied in other works, in particular \cite{HL,LS2}. More specifically, a fuzzy $d$-torus consists by definition of $d$ operators $A_1,\ldots,A_d$ which are almost unitary and almost commute, see Definition~\ref{def-Graded}. Motivated by standard models of topological insulators (see Chapter~2 in \cite{PSB}) and the work \cite{Kub}, let us associate a selfadjoint operator to the fuzzy torus:
\begin{equation}
\label{eq-GConstructIntro}
G
\;=\;
\frac{\imath}{2}\,\sum_{j=1}^{d} (A_j^*-A_j)\otimes \gamma_j+\left((d-1)\one-\frac{1}{2}\sum_{j=1}^{d}(A_j^*+A_j)\right)\otimes \gamma_{d+1}
\;,
\end{equation}
where $\gamma_1,\ldots,\gamma_{d+1}$ is an irreducible representation of the Clifford algebra with $d+1$ generators.
In Section~\ref{sec-Fuzzy}  it is shown that $G$ is gapped and hence defines an even $K$-theory class which in the case of a matrix torus of even dimension $d$ can be read out as half-signature. At the root of the construction of \eqref{eq-GConstructIntro} is a classical map from the torus $\TM^d$ to the sphere $\SM^d$ of mapping degree $1$. This map is analyzed in detail in Appendix~\ref{app-MapsExplicit}. Using variations of this map, one can construct a large set of invariants associated to the fuzzy torus, see Section~\ref{sec-Fuzzy}. In the context of Theorem~\ref{theo-Intro}, there are two fuzzy tori of matrices, namely $e^{\imath\frac{\pi}{\rho} X_1},\ldots,e^{\imath\frac{\pi}{\rho}X_d},H_\rho^\per$ and $P_\rho e^{\imath \frac{\pi}{\rho}X_1}P_\rho,\ldots,P_\rho e^{\imath \frac{\pi}{\rho} X_d}P_\rho$ where $P_\rho=\chi(H_\rho^\per<0)$. The first one is a $(d+1)$-torus consisting of $d+1$ operators, but the last operator $H_\rho^\per$ in the list is selfadjoint; such a fuzzy $(d+1)$-torus is called a graded $d$-torus (see again Definition~\ref{def-Graded}). Essentially the $G$-operator associated to the graded fuzzy $d$-torus $e^{\imath \frac{\pi}{\rho}X_1},\ldots,e^{\imath \frac{\pi}{\rho} X_d},H_\rho^\per$ is the periodic spectral localizer.  On the other hand, $P_\rho e^{\imath \frac{\pi}{\rho}X_1}P_\rho,\ldots,P_\rho e^{\imath \frac{\pi}{\rho} X_d}P_\rho$ is an un-graded fuzzy $d$-torus. The latter is the reduced out version of the former and both have the same topological content (see Proposition~\ref{Prop-Greduce}). Indeed, for $d=2$, this second fuzzy torus already played a role in \cite{ELo,ELP} and the recent work by Toniolo on quantum Hall systems \cite{Ton}. Combined with Theorem~\ref{theo-Intro} one obtains:

\begin{theo}
\label{theo-Intro2}
Let $G=G(P_\rho e^{\imath \frac{\pi}{\rho}X_1}P_\rho,\ldots,P_\rho e^{\imath \frac{\pi}{\rho} X_d}P_\rho)$ be constructed as in \eqref{eq-GConstructIntro}.  For $\rho$ sufficiently large, one has
\begin{equation}
\label{eq-EvenIntro2}
\Ch_d(P)
\;=\;
\frac{1}{2}\,\Sig(G)
\;.
\end{equation}
\end{theo}

The remainder of the paper is organized as follows. Section~\ref{sec-EvenPL} is dedicated to the proof of Theorem~\ref{theo-Intro}. Section~\ref{sec-OddPL} describes the odd dimensional version of the periodic spectral localizer. Then Section~\ref{sec-Z2PL} shows how to modify the periodic spectral localizer so that it can be used to compute $\ZM_2$-invariants in systems with real symmetries such as time-reversal and particle-hole symmetry. Finally Section~\ref{sec-Fuzzy} introduces the general notion of a fuzzy torus and shows how to extract $K$-theoretic topological invariants from it. Appendix~\ref{app-Degree} recalls the tight connection between mapping degree and Chern number, which is then applied in Appendix~\ref{app-MapsExplicit} in order to analyze the classical maps behind the index construction of the periodic spectral localizer.

\section{Periodic spectral localizer in even dimension}
\label{sec-EvenPL}

This section provides the proof of Theorem~\ref{theo-Intro}. Let $H=H^*$ be a gapped bounded selfadjoint operator on $\Hh=\ell^2(\ZM^d,\CM^L)$ of finite range $R$ which is periodic in all $d$ directions with periods ${\bf p}=(p_1,\ldots,p_d)\in\NM^d$. As above let $\rho\in\NM$ be such that $2\rho$ is an integer multiple of all these $p_j$. Then $H$ is $2\rho$ periodic in each of the $d$ directions. It is well-known that such an operator can be partially diagonalized by a Bloch-Floquet transformation $\Ff_{\rho}:\ell^2(\ZM^d,\CM^L)\to L^2(\TM^d_{\rho},\CM^{(2\rho)^d L})$ where $\TM^d_{\rho}=(\RM/4\pi{\rho}\ZM)^d$:
\begin{equation}
\label{eq-BlochFloquet}
\Ff_{\rho}\,H\,\Ff_{\rho}^*
\;=\;
\int^\oplus_{\TM^d_{\rho}} dk\;H(k)
\;.
\end{equation}
Then the periodic Hamiltonian introduced in Definition~\ref{eq-defHPer} is $H^\per_\rho=H(0)$. Actually, any fiber $H(k)$ could be used as well and Theorem~\ref{theo-Intro} remains valid. The first key observation, following directly from the direct integral representation is that 
$$
\spec(H^\per_\rho)
\,\subset\,
\spec(H)
\;.
$$ 
In particular, $H^\per_\rho$ also has a gap around the Fermi level $\mu=0$ of size at least $g=\|H^{-1}\|^{-1}$. 

\vspace{.2cm}

Next let us introduce the periodic function $\xi: \RM \to[-1,1]$ by
\begin{equation}
\label{eq-defchh}
\xi(x)\;=\;\sin(\tfrac{\pi}{2} \,x)
\;,
\end{equation}
and then set $\xi_\rho(x)=\xi(\frac{x}{\rho})$.  Due to the addition theorems, one then has
\begin{equation}
\label{eq-ChiChange}
e^{\imath \frac{\pi}{\rho}x}
\;=\;
1-2\,\xi_\rho (x)^2+\imath\,2\,\xi_\rho (x) \sqrt{1-\xi_\rho (x)^2}
\;
\end{equation}
for $x\in[-\rho,\rho]$.
The main estimates of the next lemma are folklore ({\it e.g.} \cite{BR}), but for the convenience of the reader a full proof is nevertheless provided. 

\begin{lemma}
\label{lem-Commutator}
For $j=1,\ldots,d$ and $\rho>0$, one has
\begin{equation}
\label{eq-ChhBound}
\big\|
\big[\xi _{\rho}(X_j),H\big]
\big\|
\;\leq\;
\frac{\pi}{2\rho}\, \|[X_j,H]\|
\;.
\end{equation}
Furthermore if $\rho \in \NM$ is such that $2\rho$ is an integer multiple of the periods of $H$ one has the  following commutator bounds for $H^\per_\rho$:
\begin{align}
\label{eq-ChoBound}
&
\big\|
\big[\cos(\tfrac{\pi}{\rho} X_j),H^\per_\rho\big]
\big\|
\;\leq\;
\frac{\pi}{\rho}\, \|[X_j,H]\|
\;,
\\
\label{eq-ChoBound2}
&
\big\|
\big[\sin(\tfrac{\pi}{\rho} X_j),H^\per_\rho\big]
\big\|
\;\leq\;
\frac{\pi}{\rho}\, \|[X_j,H]\|
\;,
\\
&
\big\|
\big[|\xi _{\rho}(X_j)|,H^\per_\rho\big]
\big\|^2
\;\leq\;
\frac{25\,\pi}{32\,\rho}\,\|H\|\, \|[X_j,H]\|
\;.
\label{eq-ChhBound2}
\end{align}
\end{lemma}

\noindent {\bf Proof.} Let us start out by noting that $\xi_\rho (X_j)$ is a linear combination of $e^{\imath \frac{\pi}{2\rho}\,X_j}$ and $e^{-\imath \frac{\pi}{2\rho}\,X_j}$, see \eqref{eq-defchh}. Therefore DuHamel's formula implies
$$
\|[\xi (X_j),H]\|
\;\leq\;
\frac{1}{2}
\Big(
\|[e^{\imath\frac{\pi}{2\rho}X_j},H]\|
\,+\,
\|[e^{-\imath\frac{\pi}{2\rho}X_j},H]\|
\Big)
\;\leq\;\frac{\pi}{2\rho}\,\|[X_j,H]\|
\;,
$$
which shows \eqref{eq-ChhBound}. As $\cos(\tfrac{\pi}{\rho} X_j)$ is a periodic multiplication operator one has
$$
\|
[\cos(\tfrac{\pi}{\rho} X_j),H^\per_\rho]
\|
\;\leq\;
\|
[\cos(\tfrac{\pi}{\rho} X_j),H]
\|
$$
and then \eqref{eq-ChoBound} follows from DuHamel's formula as above. Further, \eqref{eq-ChoBound2} holds by the same argument. To show \eqref{eq-ChhBound2} note that 
$$
|\xi _{\rho}(X_j)|^2
\;=\;
\frac{\one-\cos(\tfrac{\pi}{\rho} X_j)}{2}
\;.
$$
Then using the main theorem in \cite{Ped} stating that for any positve semidefinite bounded operator $T$ on $\Hh_\rho$ and any bounded operator $S$ on $\Hh_\rho$ 
$$
\|
[T^\frac{1}{2},S]
\|
\;\leq\;
\frac{5}{4}\|S\|^\frac{1}{2}\|[T,S]\|^\frac{1}{2}
\;,
$$
one obtains
\begin{align*}
\big\|
\big[|\xi _{\rho}(X_j)|,H^\per_\rho\big]
\big\|
\;\leq\;
\frac{5}{4}\|H^\per_\rho\|^\frac{1}{2}\left\|\left[\tfrac{1}{2}(\one-\cos(\tfrac{\pi}{\rho} X_j)),H^\per_\rho\right]\right\|^\frac{1}{2}
\;\leq\;
\frac{5\sqrt{\pi}}{4\sqrt{2\rho}}\|H\|^\frac{1}{2}\|[ X_j,H]\|^\frac{1}{2}
\;,
\end{align*}
where in the last step \eqref{eq-ChoBound} was used. The square of this bound is precisely \eqref{eq-ChhBound2}.
\hfill $\Box$

\vspace{.2cm}

The next result shows that the signature in \eqref{eq-EvenIntro} is well-defined. Some elements of the proof below are inspired by \cite{Kub}, others follow \cite{LS2,DSW}.

\begin{proposi}
\label{prop-gap} If all conditions of {\rm Theorem~\ref{theo-Intro}} hold, then the periodic spectral localizer satisfies the bound \eqref{eq-PLgap}. Moreover, if two parameter sets $(\eta,\rho)$ and $(\eta',\rho')$ both satisfy all conditions, then
$$
\Sig(\PL_{\eta,\rho})\;=\;\Sig(\PL_{\eta',\rho'})\;.
$$
\end{proposi}

\noindent {\bf Proof.} Let us start out with several preliminaries. To shorten notations let us denote $H^\per_{\rho}$ simply by $H_\rho$, $\sin(\frac{\pi}{\rho} X_j)$ by $s_{\rho,j}$ and $\cos(\frac{\pi}{\rho} X_j)$ by $c_{\rho,j}$. Further let us introduce the Clifford representation $\gao_1,\ldots,\gao_{d+1}$ by
$$
\gao_1=\gah_1\otimes\sigma_1\;\;,\;\ldots\,,\;\;\gao_{d-1}=\gah_{d-1}\otimes\sigma_1\,,\;\;\gao_d=\one\otimes \sigma_2\,,\;\;\gao_{d+1}=\one\otimes\sigma_3
\;,
$$
where $\sigma_1$, $\sigma_2$ and $\sigma_3$ are the Pauli matrices. Finally let us set $\tilde{H}_\rho=\frac{1}{\eta}H_\rho$. Then the periodic spectral localizer as given in \eqref{eq-PLintro} becomes
\begin{align}
\label{eq-PLhatRep}
\PL_{\eta,\rho}
&
\;=\;
\sum_{j=1}^{d} 
s_{\rho,j} \,\gao_j
\;+\;
\Big(\sum_{j=1}^{d} (1-c_{\rho,j})- \tilde{H}_\rho \Big)\gao_{d+1}
\;.
\end{align}
Next let us introduce a tapering function as in \cite{LS2,DSW} by setting $G(x)=\frac{1}{2}({\chi }(4x+3)-{{\chi} }(4x-3))$ with ${\chi}:\RM\to[-1,1]$ being be the odd non-decreasing switch function with ${\chi}(\pm x)=\pm 1$ for $x\geq 1$ given by ${\chi}(x)=x(2-|x|)$ for $x\in[-1,1]$. Then set $G_\rho(x)=G(\frac{x}{\rho})$. One finds that $G_\rho$ satisfies by construction $G_\rho(x)=1$ for $|x|\leq\frac{\rho}{2}$ and $G_\rho(x)=0$ for $|x|\geq \rho$, and, moreover, it is an even function. Furthermore, by Lemma~4 in \cite{LS1} one has $\|[G_\rho(D),H\oplus H]\|\leq \tfrac{8}{\rho}\,\|[D,H\oplus H]\|$ where $D=\sum_{j=1}^dX_j\gao_j$ (see also \cite{DSW}). As $G_\rho$ is even and $D_0=\sum_{j=1}^dX_j{\hat{\gamma}}_j$ is normal, the operator $G_\rho(D)$ is diagonal with diagonal entry $G_\rho(|D_0|)=G_\rho(|D^*_0|)$. Therefore the commutator bound can also be stated as
\begin{equation}
\label{eq-EvenLocComBound}
\|[G_\rho(|D_0|),H]\|
\;\leq\;
\frac{8}{\rho}\,\|[D_0,H]\| 
\;.
\end{equation}
Then introduce an interpolating function $G_{\rho,t}:\RM\to[0,1]$ by $G_{\rho,t}(x)=tG_{\frac{\rho}{2}}(x)+(1-t)$. 
Finally let us also set
$$
G_t
\;=\;
G_{\rho,t}(|D_0|)
\;,
\qquad
\tilde{H}_{\rho,\rho',t}\;=\;G_t\,\tilde{H}_{\rho'}\,G_t
\;,
$$
where $\rho'\in[\rho,2\rho]$ satisfies all conditions of Theorem~\ref{theo-Intro}. 
The path
\begin{equation}
\label{eq-pathPLtSig}
t\in[0,1]\mapsto
\PL_{\eta,\rho, \rho'}(t)
\;=\;
\sum_{j=1}^{d} 
s_{\rho',j} \,\gao_j
\;+\;
\Big(\sum_{j=1}^{d} (\one-c_{\rho',j})- \tilde{H}_{\rho,\rho',t} \Big)\gao_{d+1}
\end{equation}
connects $\PL_{\eta,\rho,\rho'}(0)=\PL_{\eta,\rho'}$ to an operator on $(\Hh_{\rho'}\oplus\Hh_{\rho'})\otimes \CM^{d'}$ that can be restricted to $(\Hh_{\rho}\oplus\Hh_{\rho})\otimes \CM^{d'}$ easily. After these preparation, let us now start by computing the square
\begin{align}
\PL_{\eta,\rho,\rho'}(t)^2
&\;=\;
\left(\sum_{j=1}^d s_{\rho',j}^2+\left(\sum_{j=1}^d(\one- c_{\rho',j})-\tilde{H}_{\rho, \rho', t}\right)^2\right)\otimes \one
\,+\,\sum_{1\leq j<l\leq d}[s_{\rho',j},s_{\rho',l}]\otimes \gao_j\gao_l
\nonumber
\\
&
\;\;\;\;\;\;\;\;
+\,\sum_{j=1}^d\big[s_{\rho',j},\sum_{j=1}^d(\one- c_{\rho',j})-\tilde{H}_{\rho, \rho', t}\big]\otimes \gao_j\gao_{d+1}
\nonumber
\\
&
\;\geq\;
\left(\sum_{j=1}^d s_{\rho',j}^2+\left(\sum_{j=1}^d (1-c_{\rho',j})-\tilde{H}_{\rho, \rho', t}\right)^2\right)\otimes \one
\,-\,\sum_{j=1}^d\|[s_{\rho',j},\tilde{H}_{\rho, \rho', t}]\|\one
\;.
\label{eq-Intermed}
\end{align}
because $[s_{\rho',j},s_{\rho',l} ]=[s_{\rho',j},c_{\rho',l} ]=0$ for all $j, l=1, \ldots, d$.  Hence
\begin{align*}
&\sum_{j=1}^d s_{\rho',j}^2+\left(\sum_{j=1}^d (\one-c_{\rho',j})-\tilde{H}_{\rho, \rho', t}\right)^2
\\
& \;=\;
\sum_{j=1}^ds_{\rho',j}^2+\Big(\sum_{j=1}^d (\one-c_{\rho',j})\Big)^2+(\tilde{H}_{\rho, \rho', t})^2
\,-\, G_{t}\, \sum_{j=1}^d\big((\one-c_{\rho',j})\tilde{H}_{\rho'}+\tilde{H}_{\rho'}(\one-c_{\rho',j})\big)\,G_{t}
\\
& \;=\;
\sum_{j=1}^ds_{\rho',j}^2+\Big(\sum_{j=1}^d (\one-c_{\rho',j})\Big)^2+(\tilde{H}_{\rho, \rho', t})^2
\,-\,\lambda\, G_{t} \sum_{j=1}^d\big((\one-c_{\rho',j})\hat{H}_{\rho'}+\hat{H}_{\rho'}(\one-c_{\rho',j})\big)\,G_{t}
\;,
\end{align*}
where $\lambda>0$ is a parameter to be chosen later and $\hat{H}_{\rho'}=\frac{\tilde{H}_{\rho'}}{\lambda}$.
Using $s_{\rho',j}^2+c_{\rho',j}^2=\one$ for all $j=1, \ldots, d$ one directly checks
$$
\sum_{j=1}^d s_{\rho',j}^2+\Big(\sum_{j=1}^d (1-c_{\rho',j})\Big)^2
\;\geq\;
\sum_{j=1}^d s_{\rho',j}^2+\sum_{j=1}^d (1-c_{\rho',j})^2
\;=\;
2\sum_{j=1}^d(1-c_{\rho',j})
\;.
$$
Replacing in the above gives
\begin{align*}
\sum_{j=1}^d s_{\rho',j}^2+&\left(\sum_{j=1}^d (\one-c_{\rho',j})-\tilde{H}_ {\rho,\rho',t}\right)^2 
\;\geq\;
2(1\,-\,\lambda\,G_{t}^2)\sum_{j=1}^d(\one-c_{\rho',j}) \,+\,(\tilde{H}_{\rho,\rho', t})^2
\\
&
\,+\,\lambda\,
G_{t}\sum_{j=1}^d\big((\one-c_{\rho',j})(\one-\hat{H}_{\rho'})+(\one-\hat{H}_{\rho'})(\one-c_{\rho',j})\big)G_{t}
\;.
\end{align*}
Now let us use the elementary identity
\begin{equation}
\label{eq-bound1-H}
\one-\hat{H}_{\rho'}\;=\;\frac{1}{2}(\hat{H}_{\rho'}-\one)^2+\frac{1}{2}(\one-\hat{H}_{\rho'}^2)\;,
\end{equation}
which implies that
\begin{align*}
&\sum_{j=1}^d s_{\rho',j}^2+\left(\sum_{j=1}^d(\one- c_{\rho',j})-\tilde{H}_ {\rho,\rho',t}\right)^2
\\
&
\;\geq\;
2(1\,-\,\lambda\,G_{t}^2)\sum_{j=1}^d(\one-c_{\rho',j}) \,+\,(\tilde{H}_{\rho,\rho', t})^2\\
&\;\;\;\;\;\;+\,\frac{\lambda}{2}\,G_{t}\Big(\sum_{j=1}^d((\one-c_{\rho',j})(\hat{H}_{\rho'}-\one)^2+(\hat{H}_{\rho'}-\one)^2(\one-c_{\rho',j}))\Big)G_{t}\\
&\;\;\;\;\;\;+\frac{\lambda}{2}\,G_{t}\Big(\sum_{j=1}^d((\one-c_{\rho',j})(\one-\hat{H}_{\rho'}^2)+(\one-\hat{H}_{\rho'}^2)(\one-c_{\rho',j}))\Big)G_{t}
\;.
\end{align*}
Both of the last two summands require a detailed analysis.  In order to deal with the first of them, let us use
$$
(\one-c_{\rho',j})(\hat{H}_{\rho'}-\one)^2\;=\;\big((\hat{H}_{\rho'}-\one)(\one-c_{\rho',j})+[\one-c_{\rho',j}, \hat{H}_{\rho'}-\one]\big)(\hat{H}_{\rho'}-\one)
$$
and
$$
(\hat{H}_{\rho'}-\one)^2(\one-c_{\rho',j})\;=\;(\hat{H}_{\rho'}-\one)\big((\one-c_{\rho',j})(\hat{H}_{\rho'}-\one)-[\one-c_{\rho',j}, \hat{H}_{\rho'}-\one]\big)
$$
as well as $(\hat{H}_{\rho'}-\one)(\one-c_{\rho',j})(\hat{H}_{\rho'}-\one)\geq0$. One gets
\begin{align*}
(\one-c_{\rho',j})(\hat{H}_{\rho'}- \one)^2+(\hat{H}_{\rho'}-  \one)^2(\one-c_{\rho',j})
& 
\;\geq\;-2\|[\one-c_{\rho',j}, \hat{H}_{\rho'}-\one]\|\|\hat{H}_{\rho'}-\one\|\one
\\ 
&
\;\geq\;
-2\|[c_{\rho',j}, \hat{H}_{\rho'}]\|(\|\hat{H}_{\rho'} \|+1)\one\;.
\\ 
&
\;\geq\;
-4\max\{\|\hat{H}_{\rho'}\|,1\}\,\|[c_{\rho',j}, \hat{H}_{\rho'}]\| \,\one
\\ 
&
\;\geq\;
-\frac{4\pi}{\rho'\lambda^2\eta^2}\,\max\{\|H\|,\lambda\eta\}\,\|[X_j, H]\| \,\one\;,
\end{align*}
where the last step follows from Lemma~\ref{lem-Commutator}.
This combined with $G_t^2\leq 1$ leads to 
\begin{align*}
&\sum_{j=1}^d s_{\rho',j}^2+\left(\sum_{j=1}^d(\one- c_{\rho',j})-\tilde{H}_ {\rho,\rho',t}\right)^2
\\
&
\;\geq\;
2(1\,-\,\lambda\,G_{t}^2)\sum_{j=1}^d(\one-c_{\rho',j}) \,+\,(\tilde{H}_ {\rho,\rho',t})^2
\,-\,
\frac{2 \pi}{\lambda\eta^2\rho'}\,\max\{\|H \|,\lambda\eta\}\,d\,M\,\one
\\
&\;\;\;\;\;\;+\frac{\lambda}{2}\,G_{t}\Big(\sum_{j=1}^d((\one-c_{\rho',j})(\one-\hat{H}_{\rho'}^2)+(\one-\hat{H}_{\rho'}^2)(\one-c_{\rho',j}))\Big)G_{t}
\;.
\end{align*}
For the last summand, let us use that \eqref{eq-ChiChange} implies $\frac{\one-c_{\rho',j}}{2}=|\xi _{\rho',j}|^2$ where $\xi _{\rho',j}=\sin(\frac{\pi}{2\rho'}X_j)$. Thus
\begin{align*}
(\one-c_{\rho',j}) & (\one-\hat{H}_{\rho'}^2)+(\one-\hat{H}_{\rho'}^2)(\one-c_{\rho',j})
\\
&
\;=\;
{4}\,|\xi _{\rho',j}|(\one-\hat{H}_{\rho'}^2)|\xi _{\rho',j}|
-{2}\,|\xi _{\rho',j}|[(\one-\hat{H}_{\rho'}^2),|\xi _{\rho',j}|]
-{2}\,[|\xi _{\rho',j}|,(\one-\hat{H}_{\rho'}^2)]|\xi _{\rho',j}|
\\
& 
\;\geq\;
-\,2\,\|\one-\hat{H}_{\rho'}^2\|\,(\one-c_{\rho',j})\,-\,8\frac{5\sqrt{\pi}}{\lambda^2\eta^24 \sqrt{2\rho'}}\,\|[X_j,H]\|^\frac{1}{2}\,\|H\|^\frac{3}{2}\,\one\;.
\end{align*}
Replacing in the above, one then gets
\begin{align*}
\sum_{j=1}^d  s_{\rho',j}^2+\left(\sum_{j=1}^d(\one- c_{\rho',j})-\tilde{H}_ {\rho,\rho',t}\right)^2
\;\geq\; &
\Big(2 \,- \,\big(2\lambda+\|\lambda\,\one-\lambda^{-1}\tilde{H}_{\rho'}^2\|\big)\,G_{t}^2\Big)\sum_{j=1}^d(\one-c_{\rho',j}) 
\\
&\;\,+\,(\tilde{H}_ {\rho,\rho',t})^2
\,-\,
\frac{2 \pi}{\lambda\eta^2\rho'}\,\max\{\|H \|,\lambda\eta\}\,d\,M\,\one\\
&\,-\,
\frac{5\sqrt{\pi}}{\lambda\eta^2 \sqrt{2\rho'}}\,\|H\|^\frac{3}{2}d \sqrt{M}\,\one
\;.
\end{align*}
Now let us focus on the summand $(\tilde{H}_ {\rho,\rho',t})^2$. Due to $(\tilde{H}_ {\rho,\rho',t})^2\geq \frac{g^2}{\eta^2}\,\one$ and again $G_{t}^2\leq \one$
\begin{align*}
(\tilde{H}_ {\rho,\rho',t})^2
&
\;=\;
\tfrac{1}{\eta^2}\big(G_{t}^2H_{\rho'}^2G_{t}^2\,+\,G_{t}^2H_{\rho'}[G_{t},H_{\rho'}]G_{t}\,+\,G_{t}[H_{\rho'},G_{t}]G_{t}H_{\rho'}G_{t}\big)
\\
&
\;\geq\;
\tfrac{g^2}{\eta^2}\,G_{t}^4 \,-\,\tfrac{2}{\eta^2}\|H\|\,\|[G_{t},H_{\rho'}]\|\,\one\,.
\end{align*}
The commutator can be bounded using \eqref{eq-EvenLocComBound}:
$$
\|[G_{t},H_{\rho'}]\|
\;=\;
t\|[G_{\frac{\rho}{2}}(|D_0|),H_{\rho'}]\|
\;=\;
t\|[G_{\frac{\rho}{2}}(|D_0|),H]\|
\;\leq\;
\frac{16}{\rho}\,\|[D_0,H]\|
,
$$
where $[G_{\frac{\rho}{2}}(|D_0|),H_{\rho'}]=[G_{\frac{\rho}{2}}(|D_0|),H]$ as $R+\frac{\rho}{2}\leq\rho\leq\rho'$ so that periodic boundary conditions do not interfere in the commutator. 
Replacing this, one concludes from \eqref{eq-Intermed} 
\begin{align*}
\PL_{\eta,\rho,\rho'}(t)^2
\;\geq\;&
\Big(2\,- \,\big(2\lambda+\|\lambda\,\one-\lambda^{-1}\tilde{H}_{\rho'}^2\|\big)\,G_{t}^2\Big)\sum_{j=1}^d(\one-c_{\rho',j}) \,+\,\,\frac{g^2}{\eta^2}\,G_{t}^4
\\
&\,-\,
\frac{2 \pi}{\lambda\eta^2\rho'}\,\max\{\|H \|,\lambda\eta\}\,d\,M\,\one
\,-\,
\frac{5\sqrt{\pi}}{\lambda\eta^2 \sqrt{2\rho'}}\,\|H\|^\frac{3}{2}d \sqrt{M}\,\one\\
&-\frac{32\|H\|}{\rho\eta^2}\,\|[D_0,H]\|\,\one\,-\,\frac{1}{\eta}\sum_{j=1}^d\|[s_{\rho',j},H]\|\,\one
\;.
\end{align*}
Using $\sum_{j=1}^d(1-c_{\rho',j})\leq 2d$ and also bounding $\|[s_{\rho',j},H_{\rho'}]\|\leq \frac{\pi}{\rho'} M$ by Lemma~\ref{lem-Commutator} and $\|[D_0,H]\|\leq dM$, this implies
\begin{align*}
\PL_{\eta,\rho, \rho'}(t)^2
\;\geq\;&
\Big(2\,-\, \big(2\lambda+\|\lambda\,\one-\lambda^{-1}\tilde{H}_{\rho'}^2\|\big)\,G_{t}^2\,+\,\frac{g^2}{4d\eta^2}\,G_{t}^4\Big)\sum_{j=1}^d(\one-c_{\rho',j}) \,+\,\frac{g^2}{2\eta^2}\,G_{t}^4
\\
&\,-\,
\frac{2 \pi}{\lambda\eta^2\rho'}\,\max\{\|H \|,\lambda\eta\}\,d\,M\,\one
\,-\,
\frac{5\sqrt{\pi}}{\lambda\eta^2 \sqrt{2\rho'}}\,\|H\|^\frac{3}{2}d \sqrt{M}\,\one\\
&-\frac{32\|H\|}{\rho\eta^2}\,d\,M\,-\,\frac{\pi}{\eta\rho'} \,d\,M\,\one
\;.
\end{align*}
Now the parenthesis in the first summand seen as a function of $G_{t}^2$ has a negative derivative for all $G_{\rho, t}^2\in[0,1]$ as long as 
$$
2\lambda+\|\lambda\,\one-\lambda^{-1}\tilde{H}_{\rho'}^2\|
\;\geq\;\frac{g^2}{2d\eta^2}
\;,
$$
which, after discarding the summand $\|\lambda\,\one-\lambda^{-1}\tilde{H}_{\rho'}^2\|$, actually always holds for $\lambda\geq\frac{g}{\eta}$ and $\eta\geq\frac{g}{4d}$ (which is required in Theorem~\ref{theo-Intro}). Then the minimum of the parenthesis is taken at $G_t^2=1$. Hence
\begin{align*}
\PL_{\eta,\rho,\rho'}(t)^2
\;\geq\;&
\Big(2\, -\, \big(2\lambda+\|\lambda\,\one-\lambda^{-1}\tilde{H}_{\rho'}^2\|\big)\,+\,\frac{g^2}{4d\eta^2}\Big)\sum_{j=1}^d(\one-c_{\rho',j}) \,+\,\frac{g^2}{2\eta^2}\,G_{t}^4
\\
&\,-\,
\frac{2 \pi}{\lambda\eta^2\rho'}\,\max\{\|H \|,\lambda\eta\}\,d\,M\one
\,-\,
\frac{5\sqrt{\pi}}{\lambda\eta^2 \sqrt{2\rho'}}\,\|H\|^\frac{3}{2}d \sqrt{M}\one\\
&-\frac{32\|H\|}{\rho\eta^2}\,d\,M\one\,-\,\frac{\pi}{\eta\rho'} \,d\,M\one
\\
\;\geq\;&
\Big(2\, -\, \big(2\lambda+\|\lambda\,\one-\lambda^{-1}\tilde{H}_{\rho'}^2\|\big)\,+\,\frac{g^2}{8d\eta^2}\Big)\sum_{j=1}^d(\one-c_{\rho',j}) \,+\,\frac{g^2\pi^2}{2560d\eta^2}\,\one
\\
&\,-\,
\frac{2 \pi}{\lambda\eta^2\rho'}\,\max\{\|H \|,\lambda\eta\}\,d\,M\one
\,-\,
\frac{5\sqrt{\pi}}{\lambda\eta^2 \sqrt{2\rho'}}\,\|H\|^\frac{3}{2}d \sqrt{M}\one\\
&-\frac{32\|H\|}{\rho\eta^2}\,d\,M\one\,-\,\frac{\pi}{\eta\rho'} \,d\,M\one
\;,
\end{align*}
because $\sum_{j=1}^d(1-c_{\rho',j})\geq \frac{1}{5}\sum_{j=1}^d\frac{\pi^2}{(\rho')^2}X_j^2\geq \frac{\pi^2\rho^2}{80(\rho')^2}(\one-G_t^4)\geq \frac{\pi^2}{320}(\one-G_t^4)$ where the second step holds as $(\one-G_t^4)\chi\big(\sum_{j=1}^dX_j^2\leq\frac{\rho^2}{16}\big)=0$ and the final step used the bound $\rho'\leq 2\rho$. Because $\spec(H_{\rho'})\subset\spec(H)$ 
$$
\|\lambda\,\one-\lambda^{-1}\tilde{H}_{\rho'}^2\|
\;\leq
\;\|\lambda\,\one-\lambda^{-1}\tfrac{H^2}{\eta^2}\|\;.
$$ 
Thus let us minimize $f(\lambda)=2\lambda+\|\lambda\,\one-\lambda^{-1}\frac{H^2}{\eta^2}\|$ over $\lambda\in [\frac{g}{\eta},\frac{\|H\|}{\eta}]$. By spectral calculus and elementary analysis one finds
$$
\min_{\eta\lambda\in [g,\|H\|]}\;f(\lambda)
\;=\;
\min_{\eta\lambda\in [g,\|H\|]}\;\lambda\Big(2+\max\Big\{1-\frac{g^2}{\lambda^2\eta^2}\,,\,\frac{\|H\|^2}{\lambda^2\eta^2}-1\Big\}\Big)
\;=\;f(\lambda_c)
$$
where $(\lambda_c)^2=\frac{\|H\|^2+ g^2}{2\eta^2}$. Thus
\begin{align*}
f(\lambda_c)
&
\;=\;
\frac{1}{\eta}\,\frac{g^2+3\|H\|^2}{\sqrt{2}\,\sqrt{g^2+\|H\|^2}}
\;=\;
\frac{\|H\|}{\eta}\,
\Big(
\sqrt{1-\tfrac{1}{2}(1-\tfrac{g^2}{\|H\|^2})}\;+\;\frac{1}{\sqrt{1-\tfrac{1}{2}(1-\tfrac{g^2}{\|H\|^2})}}\Big)
\\ 
& 
\;\leq\;
\frac{\|H\|}{\eta}\Big(2+\frac{1}{8}\big(1-\frac{g^2}{\|H\|^2}\big)^2\Big)
\;\leq\;
\frac{\|H\|}{\eta}\Big(2+\frac{1}{2}\big(1-\frac{g}{\|H\|}\big)^2\Big)
\;,
\end{align*}
where the first inequality follows from  $\sqrt{1+\epsilon}+\frac{1}{\sqrt{1+\epsilon}}\leq 2+\frac{1}{2}\epsilon^2$ holding for $\epsilon\in[-\frac{1}{2},1]$. Then the term in the parenthesis satisfies
$$
2- \big(2\lambda_c+\|\lambda_c\,\one-\lambda_c^{-1}\tilde{H}_{\rho'}^2\|\big)+\frac{g^2}{8d\eta^2}
\,\geq\,
2 - f(\lambda_c)+\frac{g^2}{8d\eta^2}
\,\geq\,
2 - \frac{\|H\|}{\eta}\Big(2+\frac{1}{2}\big(1-\frac{g}{\|H\|}\big)^2\Big)+\frac{g^2}{8d\eta^2}
\,\geq \,0\,
,
$$
where the last inequality is precisely the bound \eqref{eq-IntroCond2} divided by $\frac{\|H\|}{2\eta}$. Due to $\|H\|\geq\eta\lambda_c\geq g$ and the equality $\max\{\|H \|,\lambda_c\eta\}=\|H\|$ one hence deduces
\begin{align*}
\PL_{\eta,\rho,\rho'}(t)^2
\,\geq\,
\frac{g^2\pi^2}{2560d\eta^2}\,\one
&\,-\,
\frac{2 \pi}{g\eta\rho'}\|H \|\,d\,M\one
\,-\,
\frac{5\sqrt{\pi}}{g\eta \sqrt{2\rho'}}\,\|H\|^\frac{3}{2}d \sqrt{M}\one\\
&-\frac{128\|H\|}{\rho\eta g}\,d\,M\one\,-\,\frac{\pi}{\eta\rho'g} \,d\,\|H\|\,M\one\\
\,\geq\,
\frac{g^2}{300d\eta^2}\,\one
&\,-\,
\frac{128+3 \pi}{g\eta\rho}\|H \|\,d\,M\one
\,-\,
\frac{5\sqrt{\pi}}{g\eta \sqrt{2\rho'}}\,\|H\|^\frac{3}{2}d \sqrt{M}\one
\;,
\end{align*}
where $1\leq\frac{\|H\|}{g}$ and $\eta\geq\frac{g}{4}$ was used. Now $\rho$ is bounded below by \eqref{eq-IntroCond} and $\rho'$ bis bounded below by $\rho$. Then elementary numerical estimates show that 
\begin{equation}
\label{eq-Boundt}
\PL_{\eta,\rho,\rho'}(t)^2
\;\geq\;
\frac{g^2}{600\,d\,\eta^2}
\;,
\end{equation}
uniformly in $t\in[0,1]$. For $t=0$ this implies \eqref{eq-PLgap}, namely the first claim of the proposition.

\vspace{.1cm}

Now let $(\eta,\rho)$ and $(\eta',\rho')$ both satisfy all the conditions of Theorem~\ref{theo-Intro} and suppose, without restriction, that $\rho'\in[\rho,2\rho]$. Continuity of $\PL_{\eta,\rho}$ in $\eta$ together with the bound \eqref{eq-Boundt} shows that $\Sig(\PL_{\eta,\rho'})=\Sig(\PL_{\eta',\rho'})$. Hence one can assume $\eta'=\eta$. Then the above argument shows that the signature does not change along the paths $t\in[0,1]\mapsto\PL_{\eta,\rho,\rho}(t)$ and $t\in[0,1]\mapsto\PL_{\eta,\rho,\rho'}(t)$. But 
$$
\PL_{\eta,\rho,\rho'}(1)
\;=\;
\PL_{\eta,\rho,\rho'}(1)_{\Hh_\frac{\rho}{2}}\oplus (D^\per_{\rho'})_{\Hh_{\rho'}\ominus\Hh_\frac{\rho}{2}}
\;,
$$
where $D^\per_{\rho'}$ is the first summand in \eqref{eq-PLintro} with $\rho$ replaced by $\rho'$ and the lower index $\Hh_\frac{\rho}{2}$ and  $(\Hh_{\rho'}\ominus\Hh_\frac{\rho}{2})$ indicates its restriction to $(\Hh_\frac{\rho}{2}\oplus \Hh_\frac{\rho}{2})\otimes \CM^{d'}$ and $((\Hh_{\rho'}\oplus \Hh_{\rho'})\ominus (\Hh_\frac{\rho}{2}\oplus \Hh_\frac{\rho}{2}))\otimes \CM^{d'}$ respectively. As $\Sig((D^\per_{\rho'})_{\Hh_\rho'\ominus\Hh_\frac{\rho}{2}})=0$ it is sufficient to show 
$$
\Sig(\PL_{\eta,\rho,\rho'}(1)_{\Hh_\frac{\rho}{2}})
\;=\;
\Sig(\PL_{\eta,\rho,\rho}(1)_{\Hh_\frac{\rho}{2}})\;.
$$
This follows as the path
$$
t\in[0,1]\mapsto
\sum_{j=1}^{d} 
\sin(\tfrac{\pi}{t\rho+(1-t)\rho'} X_j) \,\gao_j
\,+\,
\Big(\sum_{j=1}^{d} (\one-\cos(\tfrac{\pi}{t\rho+(1-t)\rho'} X_j))-\frac{1}{\eta} G_\frac{\rho}{2}(|D_0|)H G_\frac{\rho}{2}(|D_0|) \Big)\gao_{d+1}
\,,
$$ 
of operators on $(\Hh_\frac{\rho}{2}\oplus \Hh_\frac{\rho}{2})\otimes \CM^{d'}$ entirely lies in the invertibles, which can be checked directly by an argument very similar to the one showing that the path in \eqref{eq-pathPLtSig} lies in the invertibles.
\hfill $\Box$

\vspace{.2cm}

In the proof of Proposition~\ref{prop-gap} it was shown that $\PL_{\eta,\rho}$ can be homotopically deformed into
\begin{equation}
\label{eq-PLlocalized}
\PL_{\eta,\rho,\rho}(1)
\,=\,
\sum_{j=1}^{d} 
\sin(\tfrac{\pi}{\rho}\, X_j)
\,\gao_j
\,+\,
\sum_{j=1}^{d} \big(\one-\cos(\tfrac{\pi}{\rho}\, X_j)\big)\gao_{d+1}
\,-\, \frac{1}{\eta}\, G_{\rho} H^\per_\rho G_{\rho} \,\gao_{d+1}
\end{equation}
without closing the gap, provided the conditions of Theorem~\ref{theo-Intro} hold. In particular, one has $\Sig(\PL_{\eta,\rho})=\Sig(\PL_{\eta,\rho,\rho}(1))$. Here $G_\rho=G_{\frac{\rho}{2}}(|D_0|)$ is a tapering function so that $G_{\rho} H^\per_\rho G_{\rho}$ is a tempered Hamiltonian which is localized strictly inside the volume $[-\frac{\rho}{2},\frac{\rho}{2}]^d$. In particular, the boundary conditions on the Hamiltonian are irrelevant, namely $G_{\rho} H^\per_\rho G_{\rho}=G_{\rho} H G_{\rho}$. As already stressed in Section~\ref{sec-PL}, this reflects that the signature is a local topological invariant associated  to the Hamiltonian.

\vspace{.2cm}

The next step in the proof of Theorem~\ref{theo-Intro} will be to deform the first two summands in \eqref{eq-PLlocalized}. To spell out that homotopy, it will be useful to express $\PL_{\eta,\rho,\rho}(1)$ through the function $\xi $ by means of the formula \eqref{eq-ChiChange}:
\begin{align}
\label{eq-PLhatRep2}
\PL_{\eta,\rho,\rho}(1)
&
\;=\;
2\Big(\sum_{j=1}^{d} \xi _{\rho,j}\sqrt{\one-\xi _{\rho,j}^2} \,\gao_j
\;+\;
\sum_{j=1}^{d}\xi _{\rho,j}^2\gao_{d+1}\Big) 
\;-\;\frac{1}{\eta}\, G_{\rho} H G_{\rho}
 \gao_{d+1}
\;,
\end{align}
where $\xi _{\rho,j}=\sin(\frac{\pi}{2\rho}X_j)$, see \eqref{eq-defchh}.
Then the homotopy in the parameter $s\in[0,1]$ will be given by
\begin{equation}
\label{eq-PathTwist}
\PL_{\eta,\rho}(1,s)
\;=\;
2\Big(\sum_{j=1}^{d} \xi _{\rho,j}\sqrt{\one-s^2\xi _{\rho,j}^2} \,\gao_j
\;+\;
\sum_{j=1}^{d}s\xi _{\rho,j}^2\gao_{d+1}\Big) 
\;-\; \frac{1}{\eta}\, G_{\rho} H G_{\rho} \,\gao_{d+1}
\;.
\end{equation}
Clearly $\PL_{\eta,\rho}(1,1)=\PL_{\eta,\rho, \rho}(1)$, but moreover $\PL_{\eta,\rho}(1,0)$ is essentially the spectral localizer with the damped Hamiltonian $G_{\rho} H G_{\rho}$. It was already proved in earlier works \cite{LS1,LS2,DSW} that the half-signature of the spectral localizer with this damped Hamiltonian is equal to the index pairing $\Ind(PFP+\one-P)$ with $F$ as below, which by an index theorem \cite{PSB} is  in turn equal to the Chern number. Hence a central element of the proof of Theorem~\ref{theo-Intro} consists of checking that the homotopy $s\in[0,1]\mapsto \PL_{\eta,\rho}(1,s)$ lies in the invertible matrices.

\begin{proposi}
\label{prop-Sig=Index} For $(\eta,\rho)$ satisfying the conditions of {\rm Theorem~\ref{theo-Intro}},
$$
\frac{1}{2}\,\Sig(\PL_{\eta,\rho})\;=\;\Ind(PFP+\one-P)\;,
$$
where $P=\chi(H<0)$ and  $F=D_0|D_0|^{-1}$ is the phase of $D_0=\sum_{j=1}^{d}X_j\hat{\gamma}_j$, suitably regularized at the origin.
\end{proposi}

\noindent {\bf Proof.} The main result of \cite{LS2,DSW}  states that $PFP+\one-P$ is a Fredholm operator with index that can be computed as the half-signature of the finite-volume restrictions $L_{\kappa,\rho}$ of the spectral localizer defined in \eqref{eq-SLintro}, provided that the parameters $\kappa>0$ and $\rho<\infty$ are sufficiently small and large respectively. As also the signature of the periodic spectral localizer $\PL_{\eta,\rho}$ and its damped version $\PL_{\eta,\rho,\rho}(1)$ is stable for such parameters by Proposition~\ref{prop-gap}, it hence merely has to be shown that for such parameters $\PL_{\eta,\rho, \rho}(1)$ is homotopic to $L_{\kappa,\rho}$ inside of the invertible matrices so that the signature does not change. Being able to choose $\rho$ sufficiently large considerably simplifies the proof because one can simply neglect all commutators of the type $[G_{\frac{\rho}{2}}(|D_0|),H]$ and $[\xi _{\rho,j}^2,H]$ as they are of order $\mathcal{O}(\frac{1}{\rho})$.  Here $G_{\frac{\rho}{2}}$ is the same function as used in the proof of Proposition~\ref{prop-gap}. Furthermore, it is possible to choose $\eta=\frac{5}{4}\|H\|$ because then the bound \eqref{eq-IntroCond2} is automatically satisfied.

\vspace{.1cm}

Let us start out by proving that the path $s\in[0,1]\mapsto \PL_{\eta,\rho}(1,s)$ defined in \eqref{eq-PathTwist} lies in the invertible matrices for $\rho$ sufficiently large. The proof will essentially follow the first part of the proof of Proposition~\ref{prop-gap}, namely one simply checks that $\PL_{\eta,\rho}(1,s)^2>0$ for all $s\in[0,1]$. Setting $G_\rho=G_{\frac{\rho}{2}}(|D_0|)$ and $\hat{H}=\frac{1}{\eta} H=\frac{4}{5}\frac{H}{\|H\|}$ and, moreover, discarding commutators as described above, the square of \eqref{eq-PathTwist} can be computed using
$$
\Big(\sum_{j=1}^{d} \xi _{\rho,j}\sqrt{\one-s^2\xi _{\rho,j}^2} \,\gao_j
\;+\;
\sum_{j=1}^{d}s\,\xi _{\rho,j}^2\gao_{d+1}\Big)^2 
\;\geq\;
\sum_{j=1}^{d}\xi _{\rho,j}^2
$$
and thus satisfies
\begin{align*}
&
\PL_{\eta,\rho}(1,s)^2
\;\geq\;
4\sum_{j=1}^d\xi _{\rho,j}^2+G_\rho^2\hat{H}^2G_\rho^2
-2s\Big(\sum_{j=1}^d\xi _{\rho,j}^2G_\rho \hat{H}G_\rho +G_\rho \hat{H}G_\rho \sum_{j=1}^d\xi _{\rho,j}^2\Big)
+\mathcal{O}\big(\tfrac{1}{\sqrt{\rho}}\big)
\\
&
\;=\;
4\sum_{j=1}^d\xi _{\rho,j}^2(\one-sG_\rho^2) +G_\rho^2\hat{H}^2G_\rho^2
+2sG_\rho\Big(\sum_{j=1}^d\xi _{\rho,j}^2(\one- \hat{H}) + (\one-\hat{H}) \sum_{j=1}^d\xi _{\rho,j}^2\Big)G_\rho
+\mathcal{O}\Big(\tfrac{1}{\sqrt{\rho}}\Big)
\;,
\end{align*}
with a remainder that is uniformly bounded in $s\in[0,1]$. In the first summand one can use the lower bound $\one-sG_\rho^2\geq \one-G_\rho^2$. In the second summand, let us simply use $G_\rho^2\hat{H}^2G_\rho^2 \geq \frac{16}{25}\frac{g^2}{\|H\|^2} G_\rho^4$. Finally, the last summand is non-negative up to errors $\mathcal{O}(\frac{1}{\rho})$ because $\one- \hat{H}\geq \frac{1}{5}\one$ and
$\xi _{\rho,j}^2(\one- \hat{H})=\xi _{\rho,j}(\one- \hat{H})\xi _{\rho,j}+\mathcal{O}(\frac{1}{\rho})$ by Lemma~\ref{lem-Commutator}. Hence
$$
\PL_{\eta,\rho}(1,s)^2
\;\geq\;
4\sum_{j=1}^d\xi _{\rho,j}^2(\one-G_\rho^2)
\,+\,
\frac{16}{25}\frac{g^2}{\|H\|^2} G_\rho^4
\,+\,\mathcal{O}\big(\tfrac{1}{\sqrt{\rho}}\big)
\;.
$$
Now one uses the geometric fact that $\sum_{j=1}^d\xi _{\rho,j}^2(\one-G_\rho^2)\geq \sin(\frac{\pi}{8\sqrt{d}})^2(\one-G_\rho^2)$ and the bound $1-G^2+G^4\geq \frac{3}{4}$ holding for any number $G\in[0,1]$ to conclude
\begin{align*}
\PL_{\eta,\rho}(1,s)^2
&\;\geq\;
\min\left\{\frac{16}{25}\,\frac{g^2}{\|H\|^2}, 4\sin(\tfrac{\pi}{8\sqrt{d}})^2\right\} (\one -G_\rho^2+G_\rho^4)
\,+\,\mathcal{O}\big(\tfrac{1}{\sqrt{\rho}}\big)\\
&\;\geq\;\frac{3}{4}\,\min\left\{\frac{16}{25}\,\frac{g^2}{\|H\|^2}, 4\sin(\tfrac{\pi}{8\sqrt{d}})^2\right\}\,+\,\mathcal{O}\big(\tfrac{1}{\sqrt{\rho}}\big)
\;.
\end{align*}
The next (and essentially final) step is to homotopiclly deform
$$
\PL_{\eta,\rho}(1,0)
\;=\;
\begin{pmatrix}
-G_\rho \hat{H}G_\rho& 
2 \sum_{j=1}^d \xi _{\rho,j}\gah_j^*
\\
2 \sum_{j=1}^d \xi _{\rho,j}\gah_j
&G_\rho \hat{H}G_\rho
\end{pmatrix}
$$
into
$$
\hat{L}_{\kappa,\rho}
\;=\;
\begin{pmatrix}
-G_\rho \hat{H}G_\rho&\kappa \sum_{j=1}^d X_j\gah_j^* \\
\kappa \sum_{j=1}^d X_j\gah_j &G_\rho \hat{H}G_\rho
\end{pmatrix}
\;,
$$
inside of the invertible matrices. This can readily be checked for the straight-line path from $\PL_{\eta,\rho}(1,0)$ to $\hat{L}_{\kappa,\rho}$. But due to the stability of the signature, one has $\Sig(\hat{L}_{\kappa,\rho})=\Sig({L}_{\kappa,\rho})$ for $\kappa$ sufficiently small and $\rho$ sufficiently large. But by \cite[Theorem 3]{LS2} or \cite[Theorem 10.3.1]{DSW} the index of $PFP+\one-P$ equals the half-signature of ${L}_{\kappa,\rho}$.
\hfill $\Box$

\vspace{.2cm}

\noindent {\bf Proof} of Theorem~\ref{theo-Intro}: By a well-known index theorem \cite{PSB} the Chern number $\Ch_d(P)$ is equal to the index pairing appearing in Proposition~\ref{prop-Sig=Index}, which hence directly implies the claim.
\hfill $\Box$

\begin{remark}
{\rm
As already stressed in the introduction, the quantitative aspects of the proofs in this section are far from optimal. Considerably better (but still not optimal) estimates can be obtained by working with 
$$
\PL_{\kappa,\eta,\rho}
=
\begin{pmatrix}
\sum_{j=1}^{d} \big(\one-\cos(\pi{\chi}_\kappa(X_j))\big)
&
\sum_{j=1}^{d} \sin(\pi{\chi}_\kappa(X_j))\,\gah_j^*
\\
\sum_{j=1}^{d} \sin(\pi{\chi}_\kappa(X_j))\,\gah_j
&
-\sum_{j=1}^{d} \big(\one-\cos(\pi{\chi}_\kappa(X_j))\big)
\end{pmatrix}
+
\frac{1}{\eta}
\begin{pmatrix}
-H^\per_\rho & 0 \\ 0 & H^\per_\rho
\end{pmatrix}
\;,
$$
where $\chi_\kappa(x)=\chi(\kappa x)$ is constructed from a suitable switch function $\chi$ and $\kappa>0$ is a supplementary parameter. Modifying the above proofs one can show that $\Ch_d(P)=\frac{1}{2}\,\Sig(\PL_{\kappa,\eta,\rho})$  provided that \eqref{eq-IntroCond2} holds as well as the bounds
$$
\kappa
\;\leq\;
\frac{g^3}{890\,d^2\,M\,\|H\|\,\eta}
\;,
\qquad
\rho\,\geq\,\frac{4{\sqrt{d}}}{\kappa}
\;.
$$
Note that this merely requires $\rho\geq C'\,M\|H\|\eta/g^3$ for some constant $C'$, which is considerably weaker than \eqref{eq-IntroCond}. However, the advantage of $\PL_{\eta,\rho}$ given in \eqref{eq-PLintro} over $\PL_{\kappa,\eta,\rho}$ is the simplicity of the formula as well as the fact that $\PL_{\eta,\rho}$ essentially contains no other free  parameter than the volume (as discussed in Remark~\ref{rem-EtaDiscuss}, one can safely choose $\eta\approx\|H\|$).
\hfill $\diamond$
}
\end{remark}

\section{Odd periodic spectral localizer}
\label{sec-OddPL}

In this brief section, the odd-dimensional counterpart to Theorem~\ref{theo-Intro} is described. Hence let $H$ be a periodic finite-range tight-binding Hamiltonian on the Hilbert space $\Hh=\ell^2(\ZM^d,\CM^L)$ with $d$ odd and $L$ even and with a spectral gap at $0$. On the fiber $\CM^{L}$ let $J$ be a selfadjoint unitary with eigenvalues $1$ and $-1$ of equal multiplicity $\frac{L}{2}$. The Hamiltonian is supposed to be chiral in the sense that $JHJ=-H$. This implies that it is off-diagonal in the grading of $J$:
\begin{equation}
\label{eq-ChiralH}
H
\;=\;
\begin{pmatrix}
0 & A \\ A^* & 0
\end{pmatrix}
\;,
\qquad
J
\;=\;
\begin{pmatrix}
\one & 0 \\ 0 & -\one
\end{pmatrix}
\;,
\end{equation}
where hence $A$ is an invertible short-range periodic operator  on $\ell^2(\ZM^d,\CM^{\frac{L}{2}})$. As such it has a strong invariant $\Ch_d(A)\in\ZM$ called either a (higher) winding number or also an odd Chern number \cite{PSB}. Previous results \cite{LS1,DSW} allow to compute it as the signature of the spectral localizer. Here a similar connection is established to the (odd) {\it periodic spectral localizer} which is defined to be the finite-dimensional matrix on $\Hh_\rho\otimes \CM^{d'}$ where $\Hh_\rho=\ell^2((\ZM/(2\rho\ZM))^d,\CM^L)$ given by
\begin{equation}
\label{eq-PLodd}
\PL_{\eta,\rho}
\;=\;
\begin{pmatrix}
\sum_{j=1}^d\sin(\tfrac{\pi}{\rho}\, X_j)\, \gamma_j & \sum_{j=1}^d\big(\one-\cos(\tfrac{\pi}{\rho}\, X_j)\big)
\\
\sum_{j=1}^d\big(\one-\cos(\tfrac{\pi}{\rho}\, X_j)\big) & 
-\sum_{j=1}^d\sin(\tfrac{\pi}{\rho}\, X_j)\, \gamma_j
\end{pmatrix}
\,-\,
\frac{1}{\eta}
\,H^\per_\rho
\;.
\end{equation}
Just as in \eqref{eq-PLintro}, the size $\rho$ is a multiple of the periodicities of $H$,  $H^\per_\rho$ is the Hamiltonian with periodic boundary conditions on $\Hh_\rho$ and $\gamma_1,\ldots, \gamma_d$ is an irreducible Clifford representation acting on $\CM^{d'}$. Note that $H^\per_\rho$ is again off-diagonal and its upper right entry is denoted by $A^\per_{\rho}$.

\begin{theo}
\label{theo-PLOdd}
Let $d$ be odd. Suppose that $H$ is a finite-range periodic operator on $\Hh=\ell^2(\ZM^d,\CM^L)$ of the form \eqref{eq-ChiralH}. Let $\eta$ and $\rho$ satisfy the same conditions as in {\rm Theorem~\ref{theo-Intro}}, in particular the bounds \eqref{eq-IntroCond} and \eqref{eq-IntroCond2}. Then $\PL_{\eta,\rho}$ defined in \eqref{eq-PLodd} is gapped with the bound \eqref{eq-PLgap} and the odd Chern number is
\begin{equation}
\label{eq-OddPL}
\Ch_d(A)
\;=\;
\frac{1}{2}\,\Sig(\PL_{\eta,\rho})
\;.
\end{equation}
\end{theo}

All the comments of Section~\ref{sec-PL}  transpose to the odd-dimensional case. Example~\ref{ex-Odd} in Section~\ref{sec-Fuzzy} explains that, in the case of a flat band Hamiltonian, the signature invariant in Theorem~\ref{theo-PLOdd} is in fact associated to a fuzzy torus associated to $H$.

\vspace{.2cm}

\noindent {\bf Sketch of proof} of Theorem~\ref{theo-PLOdd}.
A detailed proof will not be provided as it merely a modification of the proof of Theorem~\ref{theo-Intro}. However, let us briefly sketch the strategy of the argument. Unless differences are stressed,  the same notations as in Section~\ref{sec-EvenPL} will be used. Here $G_{t}=G_{\rho,t}(D)$ for $t\in[0,1]$ and $D=\sum_{j=1}^dX_j\, \gamma_j$. By an argument similar to the one leading to Proposition~\ref{prop-gap}, one shows that the path $t\in[0,1]\mapsto \PL_{\eta,\rho, \rho'}(t)$ for
$$
\PL_{\eta,\rho, \rho'}(t)
=
\begin{pmatrix} 
\sum_{j=1}^d\sin(\tfrac{\pi}{\rho}\, X_j)\, \gamma_j & \!\!\!\sum_{j=1}^d\big(\one-\cos(\tfrac{\pi}{\rho}\, X_j)\big)
\\
\sum_{j=1}^d\big(\one-\cos(\tfrac{\pi}{\rho}\, X_j)\big) & 
\!\!\!-\sum_{j=1}^d\sin(\tfrac{\pi}{\rho}\, X_j)\, \gamma_j
\end{pmatrix}_{\rho'}
\!\!- \frac{1}{\eta}
\begin{pmatrix} 
0 & \!\!\!\!\!\!G_{t}A^\per_{\rho'}G_{t}\\
G_{t}(A^\per_{\rho'})^*G_{t} & 0
\end{pmatrix}_{\rho'}
$$
lies in the invertibles and fulfills the bound \eqref{eq-PLgap} if $\eta$ and $\rho$ satisfy the conditions of Theorem~\ref{theo-PLOdd} and $\rho'$ fulfills $\rho\leq\rho'\leq2\rho$. Then, as in the proof of Proposition~\ref{prop-gap} one can conclude that $\Sig(\PL_{\eta,\rho})$ is independent of $\eta$ and $\rho$ in the permitted range of parameters.

\vspace{.1cm}

To show that the half-signature of the periodic spectral localizer equals the Chern number of $A$ let us use the path $s\in[0,1]\mapsto \PL_{\eta,\rho}(1,s)$ given by
\begin{equation}
\label{eq-PathTwistodd}
\PL_{\eta,\rho}(1,s)
\,=\,
\begin{pmatrix} 
2\sum_{j=1}^{d} \xi _{\rho,j}\sqrt{\one-s^2\xi _{\rho,j}^2}\, \gamma_j &2s\sum_{j=1}^{d} \xi _{\rho,j}^2-\tfrac{1}{\eta}G_t A^\per_\rho G_t
\\
2s\sum_{j=1}^{d} \xi _{\rho,j}^2-\tfrac{1}{\eta}G_t (A^\per_\rho)^*G_t& 
-2\sum_{j=1}^{d} \xi _{\rho,j}\sqrt{\one-s^2\xi _{\rho,j}^2}\, \gamma_j
\end{pmatrix}_{\rho}
\end{equation}
with $\xi _{\rho,j}=\sin(\frac{\pi}{2\rho}X_j)$ as in \eqref{eq-defchh}. As in the proof of Proposition~\ref{prop-Sig=Index}, one checks that $\PL_{\eta,\rho}(1,s)$ is invertible for $\rho$ sufficiently large. Thus $\Sig(\PL_{\eta,\rho})=\Sig(\PL_{\eta,\rho}(1,0))$. Finally, by transposing the techniques of the proof of Proposition~\ref{prop-Sig=Index},  $\PL_{\eta,\rho}(1,0)$ can be deformed inside the set of invertibles into the odd spectral localizer introduced in Section 1.4 of \cite{LS1} but with $A$ replaced by $-A$. Then as $\Ch_d(A)=\Ch_d(-A)$ Theorem~1 in \cite{LS1} allows to conclude.
\hfill $\Box$

\section{$\ZM_2$-invariants via periodic spectral localizer}
\label{sec-Z2PL}

This section addresses the real cases of the CAZ (Cartan-Altland-Zirnbauer) classification. They all impose a symmetry property on the Hamiltonian that involves a complex conjugation (real structure, denoted by an overline) on the complex Hilbert space. There are 64 such cases, stemming from an $8$-periodicity in both dimension $d$ and the CAZ classes (both routed in Bott periodicity). Only $16$ of these cases are known to lead to $\ZM_2$-valued strong invariants \cite{RSFL,GS}.  In previous works \cite{Lor,DoSB} it was shown that a real skew-adjoint version of the spectral localizer, the so-called skew localizer, can be used to compute these $\ZM_2$-indices. In this section the associated skew periodic localizer is introduced for the physically most relevant low-dimensional cases, and it is shown that the sign of its Pfaffian is connected to the $\ZM_2$-invariants.  An exhaustive treatment of all cases as in \cite{DoSB}  is not provided here.

\vspace{.2cm}

Let us now sketch the general common scheme. Like in \cite{DoSB}, the skew periodic localizer is constructed from the periodic localizer by a basis change and multiplication by $\imath$. More explicitly, in each of the relevant CAZ classes in even dimension $d$, there is a unitary $R:(\Hh_\rho \oplus \Hh_\rho)\otimes \CM^{d'}\to (\Hh_\rho\oplus \Hh_\rho)\otimes \CM^{d'}$ such that the skew periodic localizer given by 
$$
\SPL_{\eta,\rho}\;=\;\imath\, R^*\PL_{\eta,\rho} R
$$
is a real and skew-adjoint operator on $(\Hh_\rho \oplus \Hh_\rho)\otimes \CM^{d'}$. For odd $d$, the only modification is that $R: \Hh_\rho\otimes \CM^{d'}\to\Hh_\rho\otimes \CM^{d'}$ and $\SPL_{\eta,\rho}$ then acts on $\Hh_\rho\otimes \CM^{d'}$. Cleary $R$ depends on $\rho$, but as $R$ is local, this dependence is suppressed in the notation. For even $d$ and $(\eta, \rho)$ as in Theorem~\ref{theo-Intro} and for odd $d$ and $(\eta, \rho)$ as in Theorem~\ref{theo-PLOdd} this operator is invertible and therefore has a non-vanishing Pfaffian. Moreover, let $\widetilde{D}_\rho$ be a suitable perturbation of the first summand of the periodic spectral localizer \eqref{eq-PLintro} or \eqref{eq-PLodd} by a term localized at the origin such that  $\widetilde{D}_\rho$ is invertible and such that $D^{\skewu}_\rho=\imath\, R^*\widetilde{D}_\rho R$ is real and skew-adjoint. Then its Pfaffian is well-defined, does not vanish and the $\ZM_2$-index associated to the Hamiltonian $H$ is in each case proven to be given by 
\begin{equation}
\label{eq-MainZ2}
\Ind_2(T)
\;=\;
\sgn(\Pf(\SPL_{\eta,\rho}))\,\sgn(\Pf(D^{\skewu}_\rho))\,\in\,\ZM_2\;,
\end{equation}
where $\Ind_2(T)=\dim(\Ker(T))\,\mbox{\rm mod}\,2$ in the $16$ relevant cases is defined as in \cite{GS} using the Fredholm operators $T=PFP+(\one-P)$ or $T=EAE+(\one-E)$ with $F$ being the Dirac phase and $E$ the Hardy projection. Let us stress that the sign of the Pfaffian depends on the choice of basis and that for a suitable choice one can always arrange that $\sgn(\Pf(D^{\skewu}_\rho))=1$ so that the equality \eqref{eq-MainZ2} takes a more simple form. Let us also note that the index pairing and therefore also the $\ZM_2$-index does depend on the perturbation of the first summand of the periodic spectral localizer at the origin. In the following this scheme is materialized in some of the important cases by constructing the unitary $R$. Also explicit formulas for  $\SPL_{\eta,\rho}$ will be provided in these cases.

\vspace{.2cm}

\noindent {\bf Case $d=1$ for CAZ class DIII:} For $d=1$ the Dirac operator is just the position operator $D=X$. Let us add the projection onto its kernel to make it invertible, namely $\widetilde{D}= X+p_0$ where $p_0$ is the orthogonal projection onto $\Ker(D)=\text{span}(|0\rangle)$. Let $H$ be a finite-range periodic Hamiltonian on $\ell^2(\ZM, \CM^L)$ with $L$ even that is in CAZ class DIII namely that has a odd time-reversal symmetry and an even particle-hole symmetry. In a suitably chosen basis, $H$ is of the form \eqref{eq-ChiralH} with an $A\in \BM(\ell^2(\ZM, \CM^\frac{L}{2}))$ fulfilling the additional symmetry
\begin{equation}
\label{eq-SymmAd1j3}
(\imath\,\sigma_2)^*A^*\imath\,\sigma_2\;=\;\overline{A}
\end{equation}
where $\sigma_2$ denotes the second Pauli matrix acting only on the fiber. The index pairing is $T=EAE+\one-E$ where $E=\chi(\tilde{D}\geq0)$. Then set
\begin{equation}
\label{eq-Q23}
Q\;=\;
\begin{pmatrix}
0 & \imath\,\sigma_2\\
-\imath\,\sigma_2 & 0
\end{pmatrix}
\;,
\qquad
R\;=\;
\frac{1+\imath}{2}
\begin{pmatrix}
\one & \sigma_2\\
-\sigma_2 & \one
\end{pmatrix}
\;.
\end{equation}
Then $Q$ is a self-adjoint real unitary and $R$ is a particular choice for the root, namely  $R^2=Q$ (in principle one may choose other roots, but this choice leads to nice formulas below). The one-dimensional periodic spectral localizer is
$$
\PL_{\eta,\rho}
\;=\;
\begin{pmatrix}
\sin(\tfrac{\pi}{\rho} X) & \one-\cos(\tfrac{\pi}{\rho} X)\\
\one-\cos(\tfrac{\pi}{\rho} X)& -\sin(\tfrac{\pi}{\rho} X)
\end{pmatrix}
-\frac{1}{\eta}
\begin{pmatrix} 
0 & A^\per_\rho\\
(A^\per_\rho)^* & 0
\end{pmatrix}
$$
where $A^\per_\rho$ is the off-diagonal entry of $H^\per_\rho$. As $A^\per_\rho$ fulfills the same symmetry relation as $A$, namely \eqref{eq-SymmAd1j3} holds with $A$ replaced by $A^\per_\rho$, one then has $Q\overline{\PL_{\kappa,\eta,\rho}} Q=-\PL_{\kappa,\eta,\rho}$. Then set
$$
\SPL_{\eta,\rho}\,=\,\imath\, R^*\PL_{\eta,\rho} R\;,
\qquad
D^{\skewu}_\rho\,=\,\imath\, R^*\widetilde{D}_\rho R
\;,
$$
where
$$
\widetilde{D}_\rho
\;=\;
\begin{pmatrix}
\sin(\tfrac{\pi}{\rho} X)+p_0 & \one-\cos(\tfrac{\pi}{\rho} X)\\
\one-\cos(\tfrac{\pi}{\rho} X) & -\sin(\tfrac{\pi}{\rho} X)-p_0
\end{pmatrix}_\rho\;.
$$ 
Both $\SPL_{\eta,\rho}$ and $D^{\skewu}_\rho$ are bounded real and skew-adjoint operators. Explicitly one finds:
$$
\SPL_{\eta,\rho}\,=\,
\begin{pmatrix}
-\big(\one-\cos(\tfrac{\pi}{\rho} X)-\tfrac{1}{\eta}\, \Re(A^\per_\rho)\big)\imath\sigma_2 & \sin(\tfrac{\pi}{\rho} X)\imath\sigma_2+\tfrac{1}{\eta}\,\Im(A^\per_\rho)\\
\sin(\tfrac{\pi}{\rho} X)\imath\sigma_2-\tfrac{1}{\eta}\, \Im(A^\per_\rho)^* & \imath\sigma_2\big(\one-\cos(\tfrac{\pi}{\rho} X)-\tfrac{1}{\eta}\,\Re(A^\per_\rho) \big)
\end{pmatrix}
\;,
$$
where $\Re (B)=\frac{1}{2} (B+\overline{B})$  and $\Im (B)=\frac{1}{2\imath} (B-\overline{B})$ are real (note that they are different from $\Re e(B)=\frac{1}{2} (B+B^*)$  and $\Im m(B)=\frac{1}{2\imath} (B-B^*)$).

\begin{proposi}
\label{prop-skewPerLocd1j3}
For $(\eta,\rho)$ as in {\rm Theorem~\ref{theo-Intro}}, $\SPL_{\eta,\rho}$ and $D^{\skewu} $ are invertible and \eqref{eq-MainZ2} holds.
\end{proposi}

\noindent {\bf Proof.}
First of all $\PL_{\eta,\rho}$ and therefore $\SPL_{\eta,\rho}$ is invertible by Theorem~\ref{theo-PLOdd}. Therefore and as $D^{\skewu}_\rho$ is a real skew-adjoint invertible by construction, the r.h.s. of \eqref{eq-MainZ2} is well-defined. One has to show that it is independent of $\eta$ and $\rho$. For $\PL_{\eta,\rho,\rho'}(t)$, as in the proof of Theorem~\ref{theo-PLOdd},
$$
Q\,\overline{ \PL_{\eta,\rho,\rho'}(t)}\,Q
\;=\; 
- \PL_{\eta,\rho,\rho'}(t)
$$
for all $t\in[0,1]$. Therefore $\imath\, R^* \PL_{\eta,\rho,\rho'}(t)R$ is a real skew-adjoint invertible and thus its Pfaffian is well-defined, does not vanish and the sign of this Pfaffian is independent of $t$.
 Now let $(\eta,\rho)$ and $(\eta',\rho')$ both satisfy all the conditions of Theorem~\ref{theo-Intro} and suppose, without restriction, that $\rho\leq\rho'\leq 2\rho$. Continuity of $\SPL_{\eta,\rho}$ in $\eta$ allows to assume $\eta=\eta'$. By the above argument it is suffcient to show
$$
\sgn(\Pf(\imath\, R^*\PL_{\eta,\rho, \rho}(1)R))\sgn(\Pf(D^{\skewu}_\rho ))
\;=\;
\sgn(\Pf(\imath\, R^*\PL_{\eta,\rho, \rho'}(1)R))\sgn(\Pf(D^{\skewu}_{\rho'} ))
\;.
$$
But 
$$
\PL_{\eta,\rho,\rho'}(1)
\;=\;
\PL_{\eta,\rho,\rho'}(1)_{\Hh_\frac{\rho}{2}}\oplus (\widetilde{D}_{\rho'})_{\Hh_{\rho'}\ominus\Hh_\frac{\rho}{2}}
\;,
$$
where the lower index $\Hh_\frac{\rho}{2}$ indicates the restriction to $\Hh_\frac{\rho}{2}\otimes \CM^{2}$ and the lower index $(\Hh_\rho'\ominus\Hh_\frac{\rho}{2})$ indicates the restriction to $(\Hh_{\rho'}\ominus\Hh_\frac{\rho}{2})\otimes \CM^2$. Thus
\begin{align*}
\sgn(\Pf(\imath\, R^*\PL_{\eta, \rho,\rho'}(1)R))&\sgn(\Pf(D^{\skewu}_{\rho'}))
\;=\;
  \sgn(\Pf((\imath\, R^*\PL_{\eta, \rho,\rho'}(1)R)_{\Hh_\frac{\rho}{2}}))\sgn(\Pf((D^{\skewu}_{\rho'}) _{\Hh_{\rho'}\ominus\Hh_\frac{\rho}{2}}))\\&\sgn(\Pf((D^{\skewu}_{\rho'})_{\Hh_\frac{\rho}{2}} ))\sgn(\Pf((D^{\skewu}_{\rho'})_{\Hh_{\rho'}\ominus\Hh_\frac{\rho}{2}}))\\
  \;=\;&
   \sgn(\Pf((\imath\, R^*\PL_{\eta, \rho,\rho'}(1)R)_{\Hh_\frac{\rho}{2}}))\sgn(\Pf((D^{\skewu}_{\rho'})_{\Hh_\frac{\rho}{2}} ))\;.
\end{align*}
By the same argument 
\begin{align*}
\sgn(\Pf(\imath\, R^*\PL_{\eta, \rho,\rho}(1)R))&\sgn(\Pf(D^{\skewu}_{\rho}))
\;=\;
   \sgn(\Pf((\imath\, R^*\PL_{\eta, \rho,\rho}(1)R)_{\Hh_\frac{\rho}{2}}))\sgn(\Pf((D^{\skewu}_{\rho})_{\Hh_\frac{\rho}{2}} ))\;.
\end{align*}
Because the paths  $t\in[0,1]\mapsto(\imath\, R^*\PL_{\eta, \rho,t\rho+(1-t)\rho'}(1)R)_{\Hh_\frac{\rho}{2}}$ and $t\in[0,1]\mapsto(D^{\skewu}_{t\rho+(1-t)\rho'})_{\Hh_\frac{\rho}{2}}$ both lay in the real skew-adjoint invertibles 
$$
  \sgn(\Pf((\imath\, R^*\PL_{\eta, \rho,\rho'}(1)R)_{\Hh_\frac{\rho}{2}}))\;=\;  \sgn(\Pf((\imath\, R^*\PL_{\eta, \rho,\rho}(1)R)_{\Hh_\frac{\rho}{2}}))
$$
and 
$$
\sgn(\Pf((D^{\skewu}_{\rho'})_{\Hh_\frac{\rho}{2}} ))\;=\;\sgn(\Pf(D^{\skewu}_{\Hh_\frac{\rho}{2}} ))\;.
$$
for any fixed basis of  $\Hh_\frac{\rho}{2}\otimes \CM^{2}$.
This shows that the r.h.s. of \eqref{eq-MainZ2} is independent of $(\eta,\rho)$ in the permitted range of parameters.

\vspace{.1cm}

Thus it remains to show \eqref{eq-MainZ2} where $\rho$ can be chosen as large as needed. For $\PL_{\eta,\rho}(1,s)$ as in the proof of Theorem~\ref{theo-PLOdd}
$$
Q\,\overline{ \PL_{\eta,\rho}(1,s)}\,Q
\;=\; 
- \PL_{\eta,\rho}(1,s)
$$
for all $s\in[0,1]$. Thus $\imath\, R^*\PL_{\eta,\rho}(1,s)R$ is a path of real skew-adjoint invertibles. Thus for fixed and sufficiently large $\rho$, the  Pfaffain of the skew periodic localizer has the same sign as the Pfaffain of the skew localizer $\widehat{L}_{\kappa,\rho}$ for the considered index pairing introduced in Section 5.2 of \cite{DoSB} but for $-A$ instead of $A$. In the same way, one checks that the sign of the Pfaffian of $D^{\skewu}_\rho$ equals the sign of the Pfaffian of $(\imath\, R^* ((D+p_0)\oplus (D+p_0)) R)_\rho$. Then the claim follows from Theorem 26 in \cite{DoSB} as $\Ind_2(EAE+(\one-E))=\Ind_2(-EAE+(\one-E))$.
\hfill $\Box$

\vspace{.2cm}

\noindent {\bf Case $d=2$ for CAZ class AII:} 
For $d=2$ the Dirac operator is $D=X_1\gamma_1+X_2\gamma_2$. Thus its off-diagonal entry is $D_0=X_1+\imath\, X_2$. Again in order to eliminate the kernel, $D_0$ is replaced by $D_0+p_0$ with a projection $p_0$ on the origin. For sake of simplicity let us suppress this in the notations. In the present case, the index pairing is $T=PFP+\one-P$ where $F=D_0|D_0|^{-1}$ is the Dirac phase and $P=\chi(H<0)$. Then the symmetry of the Hamiltonian is $\sigma_2\overline{H}\sigma_2=H$ where $\sigma_2$ is the second Pauli matrix which commutes with $X_1$ and $X_2$. Then for $Q$ as in \eqref{eq-Q23}
 the periodic spectral localizer defined by \eqref{eq-PLintro} satisfies $Q\overline{\PL_{\eta,\rho}}Q=-\PL_{\eta,\rho}$. Finally the skew periodic localizer and $D^{\skewu}$ are again defined by
$$
\SPL_{\eta,\rho}=\imath\, R^*\PL_{\eta,\rho} R
\;,
\qquad
D^{\skewu}=\imath\, R^*D_\rho R
\;,
$$ 
for $R$ as in \eqref{eq-Q23}.
Both are real and skew-adjoint and the explicit form of the skew periodic localizer is 
\begin{align*}
\SPL_{\eta,\rho}
&
=
\begin{pmatrix}
-\imath\sigma_2 \sin(\tfrac{\pi}{\rho}X_1) &  
\!\!\!\!\!\!\!\!\!
\imath\sigma_2\sum_{j=1}^2(1-\cos(\tfrac{\pi}{\rho} X_j))+\sin(\tfrac{\pi}{\rho} X_2)
\\
\imath\sigma_2\sum_{j=1}^2(1-\cos(\tfrac{\pi}{\rho}X_j))-\sin(\tfrac{\pi}{\rho}X_2) 
& 
\!\!\!\!\!\!\!\!\!
\imath\sigma_2 \sin(\tfrac{\pi}{\rho} X_1)
\end{pmatrix}
\\
&
\;\;\;\;\;\;\;\;\;\;\;\;+
\frac{1}{\eta}
\begin{pmatrix}
\Im(H) & -\imath\sigma_2\,\Re(H) \\
-\imath\sigma_2\,\Re (H) &-\Im(H) 
\end{pmatrix}
\;.
\end{align*}
By essentially the same proof as in Proposition~\ref{prop-skewPerLocd1j3} one obtains:

\begin{proposi}
For $(\eta, \rho)$ as in {\rm Theorem~\ref{theo-Intro}}, $\SPL_{\eta,\rho}$ and $D^{\skewu}_\rho$ are invertible and \eqref{eq-MainZ2} holds.
\end{proposi}

\vspace{.2cm}

\noindent {\bf Case $d=3$ for CAZ class AII:} 
For $d=3$ the Dirac operator is $D=X_1\gamma_1+X_2\gamma_2+X_3\gamma_3$. Then let us set $\tilde{D}\;=\; D+\gamma_1p_0$. Let $H$ be a finite-range periodic Hamiltonian on $\ell^2(\ZM, \CM^L)$ that is in CAZ class AII namely that has an odd time-reversal symmetry $\sigma_2\overline{H}\sigma_2=H$. The index pairing is $T=E(\one-2P)E+\one-E$ where $E=\chi(\tilde{D}\geq0)$ and $P=\chi(H\leq0)$. The skew periodic localizer can be obtained form the even or odd periodic spectral localizer.

\vspace{.2cm}

The even periodic spectral localizer, given by \eqref{eq-PLintro}, fulfills $Q\overline{\PL_{\eta,\rho}}Q=-\PL_{\eta,\rho}$ for
\begin{equation}
\label{eq-QRj0d7}
Q
\;=\;
\begin{pmatrix}
0 & \sigma_2\gamma_2 \\ \sigma_2\gamma_2 & 0
\end{pmatrix}\;.
\end{equation}
One possible choice of $R$ is
\begin{equation}
\label{eq-Rd3j4}
R
\;=\;
\frac{1}{2}
\begin{pmatrix}
(1-\imath)\Rr & (1+\imath)\Rr \\ (1+\imath)\Rr & (1-\imath)\Rr
\end{pmatrix}
\;,
\end{equation}
where $\Rr^2=\sigma_2\gamma_2$ and $\overline{\Rr}=\Rr^*=\Rr^{-1}$. The skew periodic localizer is off-diagonal 
\begin{equation}
\label{eq-SkewLoc-d3j4}
\SPL_{\eta,\rho}
\;=\;
\begin{pmatrix}
0 &B_\rho\\
-B_\rho^* & 0
\end{pmatrix}
\;,
\end{equation}
for $B_\rho= \Rr^*(\imath\sum_{j=1}^3\sin(\tfrac{\pi}{\rho} X_j)\gamma_j-3\one+\sum_{j=1}^3\cos(\tfrac{\pi}{\rho} X_j)+\frac{1}{\eta}H^\per_\rho)\Rr $.

\begin{proposi}
\label{prop-skewPerLocd3j4}
For $( \eta,\rho)$ as in {\rm Theorem~\ref{theo-Intro}}, $\SPL_{\eta,\rho}$ and $D^{\skewu}_\rho $ are invertible and one has \eqref{eq-MainZ2}. Choosing the even periodic spectral localizer and $R$ as in \eqref{eq-Rd3j4} one obtains
$$
\Ind_2(T)
\;=\;
\sgn(\det(B_\rho))\,\sgn(\det(C_\rho))
$$
for $B_\rho$ as above and $C_\rho=\Rr^*(\imath\sum_{j=1}^3\sin(\tfrac{\pi}{\rho}X_j)\gamma_j+\gamma_1p_0-3\one+\sum_{j=1}^3\cos(\tfrac{\pi}{\rho} X_j))\Rr $.
\end{proposi}

\section{Fuzzy tori and their invariants}
\label{sec-Fuzzy}

This section develops a general theory of fuzzy tori and their invariants. Here the terminology of fuzzy geometric object is meant in the following sense: consider a classical geometric object (compact or non-compact) as a subset of an Euclidean space $\RM^d$ defined by a set of equations in the components of $x=(x_1,\ldots,x_d)\in\RM^d$; replace these coordinates or functions of them by operators in some algebra and ask the defining equations to be satisfied only approximately, namely up to errors in operator norm; then these operators are called a fuzzy geometric object. As this is a particular case of algebras defined by relation, an abstract study of the $K$-theoretic invariants of such fuzzy objects has been known for a long time  \cite{EL,ELo,ELP}. The construction of these invariants was essentially based on the replacement of the fuzzy object into classical maps from the geometric object to a sphere \cite{HL}. This leads to correct, but not very practical maps. Here we rather use relatively easy polynomial maps into the dotted Euclidean space and show that they do allow to construct the invariants, actually in a much easier manner that can be implemented numerically. The focus is only on fuzzy tori, because of their great relevance for solid state systems and hence connections to the first part of the paper. More precisely, it is shown how these abstract constructions applied to natural fuzzy tori associated to the situations analyzed in Sections~\ref{sec-PL} to \ref{sec-OddPL} directly lead to the periodic spectral localizers. 

\begin{defini}
\label{def-Graded}
Let $\Aa$ be a C$^*$-algebra of operators on a separable Hilbert space $\Hh$ and let $\Aa^\sim$ denote its unitization. Then $d$ invertible operators $A_1,\ldots,A_d\in\Aa^\sim$ form a $d$-dimensional fuzzy torus of width $\delta\in[0,1) $, or simply a fuzzy $d$-torus, if for all $j,i=1,\ldots,d$
\begin{equation}
\label{eq-deffuzzytorus2}
\|A_jA_j^*-\one\|\,\leq \delta
\;,
\qquad
\|A_j^*A_j-\one\|\,\leq \delta
\;,
\qquad
\|[A_j,A_i]\|\,\leq\,\delta
\;.
\end{equation}
If, moreover, is given a selfadjoint,
\begin{equation}
\label{eq-deffuzzytorus3}
(A_{d+1})^*\;=\;A_{d+1}
\end{equation}
such that $A_1,\ldots,A_{d+1}\in\Aa^\sim$ form a $(d+1)$-dimensional fuzzy torus of width $\delta$, then  $A_1,\ldots,A_{d+1}\in\Aa^\sim$ are said to form a graded fuzzy $d$-torus of width $\delta$.
\end{defini}

In the definition of a graded fuzzy $d$-torus, the last added selfadjoint operator $A_{d+1}$ will be viewed as in approximate grading operator which approximately commutes with the fuzzy $d$-torus given by $A_1,\ldots,A_d$.  Recall the definition of the real and imaginary part of an operator $A$:
$$
\Re e(A)\;=\; \frac{1}{2}(A+A^*)\;, \qquad
\Im m(A)\;=\; \frac{1}{2\imath}(A-A^*)\;.
$$

\begin{defini}
\label{def-GOper}
Associated to a fuzzy $d$-torus $A_1,\ldots,A_d\in\Aa^\sim$ and a subset $I\subset\{1,\ldots,d\}$, the operator $G_I=G_I(A_1,\ldots,A_d)$ is defined by
\begin{equation}
\label{eq-GDef}
G_I
\;=\;
\sum_{j=1}^{|I|} \Im m(A_{i_j})\otimes \gamma_j+\left((|I|-1)\one-\sum_{j=1}^{|I|}\Re e(A_{i_j})\right)\otimes \gamma_{|I|+1}
\;,
\end{equation}
where $I=\{i_1,\ldots,i_{|I|}\}$ and $\gamma_1,\ldots,\gamma_{d+1}$ is an irreducible selfadjoint representation of the Clifford algebra with $d+1$ generators. Furthermore, if $A_1,\ldots,A_{d+1}\in\Aa^\sim$ form a graded fuzzy $d$-torus, then an operator $\widehat{G}_I=\widehat{G}_I(A_1,\ldots,A_{d+1})$ is introduced by
\begin{equation}
\label{eq-GhatDef}
\widehat{G}_I
\;=\;
\sum_{j=1}^{|I|} \Im m(A_{i_j})\otimes \gamma_j+\left(|I|\one-\sum_{j=1}^{|I|} \Re e(A_{i_j})-A_{d+1}\right)\otimes \gamma_{|I|+1}
\;.
\end{equation}
\end{defini}

By construction, $G_I=G_I^*$ and $\widehat{G}^*_I=\widehat{G}_I$ are selfadjoint operators on $\Hh\otimes\CM^{d'}$ for some $d'$. The results below show that $G_I$ and $\widehat{G}_I$ are invertible for sufficiently small $\delta$, so that their positive spectral projections fix $K$-theory classes which for matrices can simply be read out via the signature. Underlying the construction in  \eqref{eq-GDef} are certain maps $g_{|I|,|I|-1}:\TM^d\to\RM^{d+1}\setminus\{0\}$ which for even $|I|$ are analyzed in detail in Appendix~\ref{app-MapsExplicit}. In particular, it is shown that the normalized maps $g_{|I|,|I|-1}/\|g_{|I|,|I|-1}\|:\TM^d\to\SM^{d}$ have a mapping degree equal to $1$. One then gets $G_I:\TM^d\to\CM^{d'\times d'}$ by multiplying the coefficients with an irreducible representation of $|I|+1$  Clifford generators:
$$
G_I(e^{\imath \theta})
\;=\;
\sum_{j=1}^{|I|} \sin(\theta_{i_j})\otimes \gamma_j+\left((|I|-1)\one-\sum_{j=1}^{|I|}\cos(\theta_{i_j})\right)\otimes \gamma_{|I|+1}
\;.
$$
As $G_I(e^{\imath \theta})$ remains gapped, $\Sig(G_I(e^{\imath \theta}))$ is independent of $\theta$ and  $\Sig(G_I(e^{\imath \theta}))=\Sig(G_I(1))=0$. The same holds for fuzzy tori composed of commuting unitary matrices:

\begin{proposi}
\label{prop-CommutingTori}
If a fuzzy $d$-torus consists of commuting unitary matrices $A_1,\ldots,A_d$, one has $\Sig(G_I)=0$ and $\Sig(\widehat{G}_I)=0$ for all $I$.  
\end{proposi}

\noindent {\bf Proof.} As the matrices can be simultaneously diagonalized, the above argument applies to all common eigenspaces.
\hfill $\Box$

\vspace{.2cm}

For non-commuting (but almost commuting) matrices, the signatures of $G_I$ and $\widehat{G}_I$ can be different from zero though, see the examples below. Hence these signatures allow to distinguish different homotopy classes of fuzzy $d$-tori.  As shown in Proposition~\ref{Prop-Greduce} below, the definition \eqref{eq-GhatDef} of $\widehat{G}_I$ essentially reduces to the same map.  Furthermore, one has the following elementary fact:

\begin{lemma}
\label{lem-GGrading}
For odd $|I|$, there exists a further Clifford generator $\Gamma=\gamma_{|I|+2}$ on the same representation space for which $\Gamma G_I\Gamma=-G_I$ and $\Gamma \widehat{G}_I\Gamma=-\widehat{G}_I$.
\end{lemma}

\begin{remark}
{\rm 
The formulas \eqref{eq-GDef} and \eqref{eq-GhatDef} look very much alike, but there is nevertheless a crucial difference that will be explained now. Given a graded  fuzzy $d$-torus $A_1,\ldots,A_{d+1}\in\Aa^\sim$, one can, of course, view it by definition as a $(d+1)$-dimensional fuzzy torus and hence associate the operators  $G_I=G_I(A_1,\ldots,A_{d+1})$. It requires the use of an irreducible selfadjoint representation of the Clifford algebra with $|I|+1$ generators even though, for $i_{|I|}=d+1$ and due to $\Im m(A_{d+1})=0$, the generator $\gamma_{|I|}$ does not appear in the formula. Therefore, the operator $G_I$ satisfies the chirality relation $\gamma_{|I|}G_I\gamma_{|I|}=-G_I$, no matter whether $|I|$ is even or odd. For even $|I|$ so that $|I|+1$ is odd, it is not possible to add a further Clifford generator on the representation space. On the other hand, if $|I|$ is odd and $|I|+1$ is even, there then {\it does} exist an extra generator $\gamma_{|I|+2}$. Choosing the representation such that
$$
\gamma_1\,=\,\hat{\gamma}_1\otimes\sigma_1
\;,\;
\ldots
\;,\;
\gamma_{|I|-1}\,=\,\hat{\gamma}_{|I|-1}\otimes\sigma_1
\;,\;\;\;
\gamma_{|I|+1}\,=\,\hat{\gamma}_{|I|}\otimes\sigma_1
\;,\;\;\;
\gamma_{|I|}\,=\,\one\otimes \sigma_2
\;,\;\;\;
\gamma_{|I|+2}\,=\,\one\otimes \sigma_3
\;,
$$
where $\sigma_1$, $\sigma_2$ and $\sigma_3$ are the Pauli matrices and $\hat{\gamma}_{1},\ldots,\hat{\gamma}_{|I|}$ is an irreducible selfadjoint representation of the Clifford algebra with $|I|$ generators, one then has the two chirality relations
$$
\sigma_2 G_I \sigma_2
\;=\;-G_I
\;,
\qquad
\sigma_3 G_I \sigma_3
\;=\;-G_I
\;.
$$
By an elementary argument with $2\times 2$ matrices, the second relation implies that $G_I$ is off-diagonal in the grading of the Pauli matrices, and the first relation that the off-diagonal entry is selfadjoint. Actually, setting $\hat{I}=I\setminus\{d+1\}$ and comparing with the definition of $\widehat{G}_{\hat{I}}$ written with the $\hat{\gamma}_j$ instead of the $\gamma_j$, one finds
\begin{equation}
\label{eq-OddReduce}
G_I
\;=\;
\begin{pmatrix}
0 & \widehat{G}_{\hat{I}} \\ \widehat{G}_{\hat{I}} & 0
\end{pmatrix}
\;,
\qquad
|I|\;\mbox{ odd}
\;.
\end{equation}
Hence $\widehat{G}_{\hat{I}}$ is the reduced-out form of $G_I$. Note that the spectra satisfy $\spec(G_I)=\spec(\widehat{G}_{\hat{I}})\cup(-\spec(\widehat{G}_{\hat{I}}))$. In particular, the spectrum of $G_I$ is always symmetric, while that of $\widehat{G}_{\hat{I}}$ may have a spectral asymmetry. 
\hfill $\diamond$
}
\end{remark}

Before starting with the analysis of the gap of $G_I$ and $\widehat{G}_I$  and their $K$-theoretic interpretations, let us provide several examples of fuzzy tori.

\begin{example}
\label{eq-NCtorus}
{\rm 
Let $\theta\in\RM^{d\times d}$ be an anti-symmetric matrix and $\Aa_\theta=C^*(U_1,\ldots,U_d)$ be the $d$-dimensional rotation algebra generated by $d$ unitaries $U_1,\ldots,U_d$ satisfying $U_iU_j=e^{\imath \theta_{i,j}}U_jU_i$. One has the bound $\|[U_i,U_j]\|=(2(1-\cos(\theta_{i,j})))^\frac{1}{2}$. Hence if $|\theta_{i,j}|\leq \delta$, then $\|[U_i,U_j]\|\leq\delta$. If this holds for all $i\not=j$, then $U_1,\ldots,U_d$ form a fuzzy $d$-torus of width $\delta$. 
}
\hfill $\diamond$
\end{example}

\begin{example}
\label{ex-Odd0}
{\rm 
Let $A\in C^1(\TM^d,\CM^{L\times L})$ be a continuously differentiable function from the classical $d$-torus to the invertible $L\times L$-matrices. In applications to solid state physics, this $A$ appears as the off-diagonal entry of a periodic Hamiltonian with chiral symmetry \cite{PSB}. This is viewed as a fiberwise multiplication operator on $L^2(\TM^d,\CM^L)$. It is supposed to lead to small norms $\|A^*A-\one\|$ and $\|AA^*-\one\|$, namely is almost unitary. It is well-known that $A$ has odd Chern numbers (also called higher winding numbers) given by
$$
\Ch_d(A)
\;=\;
\frac{(\frac{1}{2}(d-1))!}{d !}
\;
\left(\frac{\imath}{2\pi}\right)^{\frac{d+1}{2}}\,
\int_{\TM^d}\Tr\left(\big ( A^{-1} {\bf d} A \big )^d \right)
\;.
$$
Let $\imath \partial_j$ for $j=1,\ldots,d$ be the (selfadjoint) coordinate vector fields on the torus. Then $e^{\imath \pi \chi _\kappa(\imath\partial_j)}$ is defined by functional calculus from a scaled smooth switch function $\chi_\kappa(x)=\chi(\kappa x)$. For $\kappa$ sufficiently small, an argument similar to the one leading to Lemma~\ref{lem-Commutator} implies that 
$$
e^{\imath \pi \chi _\kappa(\imath\partial_1)}\;,\ldots,e^{\imath \pi \chi _\kappa(\imath\partial_d)}\,,A
$$
form a fuzzy $(d+1)$-torus in the algebra of bounded operators on $L^2(\TM^d,\CM^L)$. Its width can be determined from $\kappa$ and the above two norms. Note that, due to $\chi _\kappa(\imath\partial_j)=\chi (\kappa\imath\partial_j)$, $\kappa$ plays the role of Planck's constant here. Also let us stress that associated to a classical $d$-torus $\TM^d$ is a fuzzy $(d+1)$-torus. Hence there there is a natural dimensional shift here. There is an associated self-adjoint operator $G=G_{\{1,\ldots,d+1\}}$. If one chooses the Clifford representation
$$
\gamma_1\otimes\sigma_3\,,\ldots,\;\gamma_d\otimes\sigma_3\,,\;\one\otimes\sigma_2\,,\;\one\otimes\sigma_1\;,
$$
it is given by
\begin{equation}
\label{eq-GClassTorus}
G
\,=\,
\begin{pmatrix}
\sum_{j=1}^d\sin(\pi\chi _\kappa (\imath\partial_j))\, \gamma_j & d-\sum_{j=1}^d\cos(\pi\chi _\kappa (\imath\partial_j))
\\
d-\sum_{j=1}^d\cos(\pi\chi _\kappa (\imath\partial_j)) & 
-\sum_{j=1}^d\sin(\pi\chi _\kappa (\imath\partial_j))\, \gamma_j
\end{pmatrix}
\,-\,
\begin{pmatrix}
0 & A \\ A^* & 0
\end{pmatrix}
\,.
\end{equation}
Note that if $d$ is even, then the Clifford representation $\gamma_1,\ldots,\gamma_d$ admits another generator $\gamma_{d+1}$ and then $G$ is odd w.r.t. $\gamma_{d+1}\otimes\sigma_3$. 
}
\hfill $\diamond$
\end{example}

\begin{example}
\label{ex-Odd}
{\rm 
Upon Fourier transform $\Ff:\ell^2(\ZM^d,\CM^L)\to L^2(\TM^d,\CM^L)$, Example~\ref{ex-Odd0} essentially becomes the situation described in Section~\ref{sec-OddPL} because $\Ff^* \imath\partial_j\Ff=X_j$ and the finite range condition in Section~\ref{sec-OddPL} corresponds to a finite frequency condition. Then $\Ff^* A\Ff$ is a $1$-periodic operator on $\ell^2(\ZM^d,\CM^L)$ which for sake of simplicity is simply denoted by $A$ again. Then $e^{\imath \pi \chi _\kappa(X_1)},\ldots,e^{\imath \pi \chi _\kappa(X_d)},A$ form a fuzzy $(d+1)$-torus in the algebra of bounded operators on $\ell^2(\ZM^d,\CM^L)$, actually rather the much smaller algebra generated by the algebra $\Aa_1$ of $1$-periodic short-range operators and the algebra $\Kk$ of compact operators on $\ell^2(\ZM^d,\CM^L)$ (as a vector space, this algebra is $\Aa_1\oplus\Kk$, but the multiplication is not fiberwise).  The associated operator $G=G_{\{1,\ldots,d+1\}}$ is, as in \eqref{eq-GClassTorus},
$$
G
\,=\,
\begin{pmatrix}
\sum_{j=1}^d\sin(\pi\chi _\kappa (X_j))\, \gamma_j & d-\sum_{j=1}^d\cos(\pi\chi _\kappa (X_j))
\\
d-\sum_{j=1}^d\cos(\pi\chi _\kappa (X_j)) & 
-\sum_{j=1}^d\sin(\pi\chi _\kappa (X_j))\, \gamma_j
\end{pmatrix}
\,-\,
\begin{pmatrix}
0 & A \\ A^* & 0
\end{pmatrix}
\,.
$$
In this formula, one can now let $A$ be a $2\rho$ periodic operator. Furthermore, choosing $\chi(x)=x$ for $|x|\leq 1$ and $\chi(x)=\sgn(x)$ for $|x|\geq 1$ as well as $\kappa=\frac{1}{\rho}$, one then obtains an operator $G$ with a restriction $G_\rho$ to $\Hh_\rho=\ell^2((\ZM/(2\rho\ZM))^d,\CM^L)$ coinciding with the odd periodic spectral localizer with $\eta=1$, provided one replaces $A$ by $A_\rho$ with periodic boundary conditions. The spectral asymmetry of $G$ entirely results from the spectral asymmetry of this finite-dimensional piece $G_\rho$ (as can readily be shown for $\rho$ sufficiently large and the arguments in Section~\ref{sec-EvenPL} that modifying the boundary conditions does not alter the signature either) and can hence be measured by the half-signature of the odd periodic spectral localizer. The fuzzy torus (of square matrices of size $(2\rho)^dL$) leading to $G_\rho$ is given by $e^{\imath \frac{\pi}{\rho} X_1},\ldots,e^{\imath \frac{\pi}{\rho} X_d},A_\rho$.
Actually, the framework can further be extended to operators $A$ from the algebra $\Aa$ of covariant operators as defined in \cite{Bel,PSB}. Then the fuzzy torus lies in the algebra $\Aa\oplus\Kk$. 
}
\hfill $\diamond$
\end{example}

\begin{example}
\label{ex-Even}
{\rm 
Suppose that $H=H^*$ is an invertible finite-range operator on $\ell^2(\ZM^d,\CM^L)$.  Then the operators
$$
e^{\imath \pi \chi _\kappa(X_1)}\,,\ldots,e^{\imath \pi \chi _\kappa(X_d)}\,,\;H
\;.
$$
form a graded fuzzy $d$-torus. Associated is therefore an operator $\widehat{G}=\widehat{G}_{\{1,\ldots,d\}}$. If the irreducible representation of the Clifford algebra with $d+1$ generators is chosen to be
$$
\gamma_1\otimes\sigma_1\,,\ldots\,,\,\gamma_{d-1}\otimes\sigma_1\,,\,\one\otimes \sigma_2\,,\,\one\otimes\sigma_3
$$
where $\gamma_1,\ldots,\gamma_{d-1}$ is an irreducible representation of the Clifford algebra with $d-1$ generators (the tensor products will be dropped in the following), then one finds
\begin{align*}
\widehat{G}
&
\;=\;
\sum_{j=1}^{d-1} \Im m(e^{\imath \pi \chi _\kappa(X_j)}) \,\gamma_j\,\sigma_1+
\Im m(e^{\imath \pi \chi _\kappa(X_d)})\sigma_2
+\Big(d\,\one-\sum_{j=1}^{d}\Re e(e^{\imath \pi \chi _\kappa(X_1)})-H\Big)\sigma_3
\\
&
\;=\;
\begin{pmatrix}
-H+d\,\one-\sum_{j=1}^{d} \cos(\pi\chi _\kappa(X_j))
&
\sum_{j=1}^{d-1} \sin(\pi\chi _\kappa(X_j))\,\gamma_j-\imath \sin(\pi\chi _\kappa(X_d))
\\
\sum_{j=1}^{d-1} \sin(\pi\chi _\kappa(X_j))\,\gamma_j+\imath \sin(\pi\chi _\kappa(X_d))
&
H-d\,\one+\sum_{j=1}^{d} \cos(\pi\chi _\kappa(X_j))
\end{pmatrix}
\;.
\end{align*}
Replacing the $\gamma_j$ by $\hat{\gamma}_j$ and proceeding as in Example~\ref{ex-Odd}, one recovers the even periodic spectral localizer with $\eta=1$ as the finite-volume restriction of $\widehat{G}$. In connection with this example, let us also point out that Proposition~\ref{Prop-Greduce} below shows how to construct an (ungraded) fuzzy $d$-torus associated to the above graded fuzzy torus. It is given by $Pe^{\imath \pi \chi _\kappa(X_1)}P,\ldots,Pe^{\imath \pi \chi _\kappa(X_d)}P$ where $P=\chi(H>0)$. This fuzzy $d$-torus already appeared in \cite{Ton}, and Proposition~\ref{Prop-Greduce} shows that its associated $G$-operators $G^P_I$ are homotopic to the $\widehat{G}_I$ and therefore have the same signature invariants.
}
\hfill $\diamond$
\end{example}

\begin{example}
\label{ex-GeneralTorus}
{\rm 
Suppose given a $C^*$-dynamical system $(\Aa,\RM^d,\alpha)$ consisting of a $C^*$-algebra (for simplicity given as subalgebra of the bounded operators on some Hilbert space $\Hh$) and a continuous group action $\alpha$ of $\RM^d$ on $\Aa$. The action is implemented (in a unique manner, up to isomorphisms) by a strongly continuous unitary group action $U$ on $\Hh$, namely $\alpha_x(A)=U_x^*AU_x$ for $A\in\Aa$ and $x\in\RM^d$. If $e_1,\ldots,e_d$ is a basis of $\RM^d$, then $t\in\RM\mapsto U_{te_j}$ is a strongly continuous one-parameter group with generator $X_j$. Then define unitaries $e^{\imath \chi_\kappa(X_j)}$ by spectral calculus. Further be given an invertible operator $A\in\Aa$ fixing a class in $K_1(\Aa)$. This operator is supposed to be sufficiently smooth w.r.t. the action $\alpha$  (see \cite{SBS} for a detailed description of conditions that assure the existence of semi-finite index parings in the presence of an $\alpha$-invariant tracial state on $\Aa$). One then has an associated fuzzy $(d+1)$-torus $e^{\imath \chi_\kappa(X_1)},\ldots,e^{\imath \chi_\kappa(X_d)},A$ of width converging to $0$ as $\kappa\to 0$ (no detailed proof of this fact is provided here, as it readily follows from the techniques of Lemma~\ref{lem-Commutator} and \cite{DSW}). Similarly, given a sufficiently smooth gapped selfadjoint operator $H\in\Aa$ specifying a class in $K_0(\Aa)$, one has a graded fuzzy $d$-torus $e^{\imath \chi_\kappa(X_1)},\ldots,e^{\imath \chi_\kappa(X_d)},H$. From both of these tori, one can build lower-dimensional tori associated to choices of a subset $I\subset\{1,\ldots,d\}$. Note that these constructions are generalizations of the Examples~\ref{ex-Odd} and \ref{ex-Even}.
}
\hfill $\diamond$
\end{example}

The next two results give simple criteria on the width $\delta$ assuring that $G$-operators are invertible. We will focus on the case $G=G_{\{1,\ldots,d\}}$ simply because the $G_I$ with $|I|<d$ are associated to a fuzzy $|I|$-torus so that the below results cover this case as well.  As the estimate of the gap is more simple and transparent,  let us first restrict to a fuzzy $d$-torus given by unitary instead of invertible operators. Note that this case is sufficient for Example~\ref{eq-NCtorus}.

\begin{proposi}
\label{Prop-gapG}
For unitaries $U_1, \ldots, U_d\in \UM(\Hh)$, the operator $G=G_{\{1,\ldots,d\}}(U_1,\ldots,U_d)$ satisfies 
\begin{equation}
\label{eq-gapG}
G^2
\;\geq\;
\left(1-7\sum_{1\leq j<i\leq d}\|[U_j,U_i]\|\right)\one\;.
\end{equation}
In particular, $G$ is invertible if the unitaries $U_1, \ldots, U_d\in \UM(\Hh)$ form a fuzzy $d$-torus of width $\delta$ satisfying
$1-\binom{d}{2}7\delta>0$.
\end{proposi}

\noindent {\bf Proof.}
The argument is essentially identical to the one leading to Proposition 4.3 in \cite{Kub}. Using the Clifford relations, one finds
\begin{align*}
G^2
\;=\;&
\left(\sum_{j=1}^d \Im m(U_j)^2+\left((d-1)\one-\sum_{j=1}^d \Re e(U_j)\right)^2\right)\otimes \one
+\,\sum_{1\leq j<i\leq d}[\Im m(U_j),\Im m(U_i)]\otimes \gamma_j\gamma_i\\
&\;\;+\,\sum_{j=1}^d[\Im m(U_j),(d-1)\one-\sum_{i=1}^d \Re e(U_i)]\otimes \gamma_j\gamma_{d+1}\;.
\end{align*}
Using that 
$$
\|[\Im m(U_j),\Im m(U_i)]\|\;\leq\;\|[U_j,U_i]\|\;,
\qquad
\|[\Im m(U_j),\Re e(U_i)]\|\;\leq\;\|[U_j,U_i]\|
$$
and that $\gamma_j$ is unitary for all $j$ one obtains
\begin{align}
G^2
\;\geq\;
&
\label{eq-boundComReIm}
\left(\sum_{j=1}^d \Im m(U_j)^2+\left((d-1)\one-\sum_{j=1}^d \Re e(U_j)\right)^2\right)\otimes \one\,-\,3\sum_{1\leq j<i\leq d}\|[U_j,U_i]\|\one\;.
\end{align}
Moreover,
\begin{align*}
\sum_{j=1}^d &\Im m(U_j)^2+\left((d-1)\one-\sum_{j=1}^d \Re e(U_j)\right)^2\\
\;=\;&
\sum_{j=1}^d \Im m(U_j)^2+(d-1)^2\one-2(d-1)\sum_{j=1}^d \Re e(U_j)+\left(\sum_{j=1}^d \Re e(U_j)\right)^2\\
\;=\;&
d\one+(d-1)^2\one-2(d-1)\sum_{j=1}^d \Re e(U_j)+\sum_{1\leq j<i\leq d}^d (\Re e(U_j)\Re e(U_i)+\Re e(U_i)\Re e(U_j))\;,
\end{align*}
where the last step follows from $\Im m(U_j)^2+\Re e(U_j)^2=\one$. This simplifies to
\begin{align}
\nonumber
\sum_{j=1}^d &\Im m(U_j)^2+\left((d-1)\one-\sum_{j=1}^d \Re e(U_j)\right)^2
\\
\label{eq-boundSumRe}
& \;=\;
\one+\sum_{1\leq j<i\leq d}\big((\one-\Re e(U_j))(\one-\Re e(U_i))+(\one-\Re e(U_i))(\one-\Re e(U_j))\big)\;.
\end{align}
One directly checks that $\one-\Re e(U_j)=\frac{1}{2}(U_j-\one)(U_j-\one)^*$ and therefore
\begin{align*}
& \sum_{j=1}^d  \Im m(U_j)^2+\left((d-1)\one-\sum_{j=1}^d \Re e(U_j)\right)^2
\\
& \;=\;
\one+\frac{1}{4}\sum_{1\leq j<i\leq d} \Big((U_j-\one)(U_j-\one)^*(U_i-\one)(U_i-\one)^* +(U_i-\one)(U_i-\one)^*(U_j-\one)(U_j-\one)^*\Big)
\\
& \;=\;
\one+\frac{1}{4}\sum_{1\leq j<i\leq d}\!\! \Big(
(U_j-\one)(U_i-\one)(U_i-\one)^*(U_j-\one)^*+(U_j-\one)[(U_j-\one)^*,(U_i-\one)](U_i-\one)^*
\\
& \hspace{2.0cm} +(U_j-\one)(U_i-\one)[(U_j-\one)^*,(U_i-\one)^*]
+(U_i-\one)(U_j-\one)(U_j-\one)^*(U_i-\one)^*
\\
& \hspace{2.0cm} +(U_i-\one)[(U_i-\one)^*,(U_j-\one)](U_j-\one)^*+(U_i-\one)(U_j-\one)[(U_i-\one)^*,(U_j-\one)^*]\Big)
.
\end{align*}
Using that the first and fourth summand are non-negative and
%
$\|(U_j-\one)\|\leq 2$, $\|(U_i-\one)\|\leq 2$, one obtains
\begin{align*}
\sum_{j=1}^d \Im m(U_j)^2+&\left((d-1)\one-\sum_{j=1}^d \Re e(U_j)\right)^2
\;\geq\;
\big(1-4\sum_{1\leq j<i\leq d}\|[U_j,U_i]\|\big)\one\;.
\end{align*}
Combining with \eqref{eq-boundComReIm} the claim \eqref{eq-gapG} follows.
\hfill $\Box$

\begin{example}
\label{eq-NCtorus2}
{\rm 
Let us continue with Example~\ref{eq-NCtorus} of the non-commutative torus. If $|\theta_{i,j}|\leq \delta$ for all $i\not=j$ and $\delta<\frac{2}{7}d(d-1)$, then $G_I=G_I(U_1,\ldots,U_d)$ is gapped by Proposition~\ref{Prop-gapG}.  If $\frac{1}{2\pi}\,\theta$ consists only of rational numbers, it is well-known that the unitaries $U_1,\ldots,U_d$ can be chosen to be finite-dimensional matrices. In this case also $G_I$ is a finite-dimensional matrix and one can hence define the invariants $\nu_I=\frac{1}{2}\,\Sig(G_I)$ which are integer-valued because the representation space of the Clifford algebra is even-dimensional so that also the selfadjoint matrix $G_I$ acts on an even-dimensional vector space. Let us now focus on the case $d=2$ and $I=\{1,2\}$. Then $\theta$ is a scalar which is supposed to be $\theta=\frac{2\pi}{N}$. Choosing the Clifford representation to be the standard Pauli matrices, the associated operator $G=G_{\{1,2\}}(U_1,U_2)$ is then
$$
G
\;=\;
\begin{pmatrix}
\one-\Re e(U_1) -\Re e(U_2) & \Im m(U_1) -\imath\,\Im m(U_2) \\
 \Im m(U_1) +\imath\,\Im m(U_2)  & -\one+\Re e(U_1) +\Re e(U_2)
\end{pmatrix}
\;.
$$
By Proposition~\ref{Prop-gapG}, $G$ is gapped provided that $N\geq 14\,\pi$. In reality, the gap is already open for $N$ much smaller.  Associated is then the invariant $\frac{1}{2}\,\Sig(G(U_1,U_2))\in\ZM$. Due to \cite[Proposition~5.1]{ELo} and \cite[Theorem~6.15]{ELP} it is known that this integer is equal to the winding number of the path $t\in[0,1] \mapsto\det(tU_1U_2-(1-t)U_2U_1)$.
}
\hfill $\diamond$
\end{example}

\begin{example}
\label{ex-Odd0bis}
{\rm 
This example is a continuation of Example~\ref{ex-Odd0}, albeit with a function $A=U\in C^1(\TM^d,\mbox{\rm U}(L))$ with values in the unitary matrices. For such a function, Proposition~\ref{Prop-gapG} applies. Therefore the $G$-operator given in \eqref{eq-GClassTorus} is gapped. It is still an operator on an infinite-dimensional Hilbert space and has no finite-dimensional invariant subspaces (even if $U$ only contains a finite number of frequencies). Nevertheless, the operator has a spectral asymmetry which can be extracted by projecting $G$ down to frequencies of modulus less then $\rho$. If $G_\rho$ denotes this restriction, then the proof of Theorem~\ref{theo-PLOdd} shows that $\Ch_d(U)=\frac{1}{2}\Sig(G_\rho)$ for $\rho$ sufficiently large. Let us note that the spectrum of the operator $G$ in infinite volume is contained in two intervals of size $\kappa$ around  $-1$ and $1$, because  $G^2-\one$ is bounded by a constant time $\kappa$. In particular, $G$ does not have a compact resolvent so that it is not possible to define an $\eta$-invariant, other than for the spectral localizer \cite{LS1,DSW}. Furthermore, let us note that for $d$ even, the symmetry $\gamma_{d+1}\otimes\sigma_3G_\rho\gamma_{d+1}\otimes\sigma_3 =-G_\rho$ implies that $\Sig(G_\rho)=0$. However, for $d$ even, one can choose $I$ of odd cardinality $|I|<d$ and then the spectral asymmetry of the associated operators $G_I$ determines the odd Chern numbers of lower degree.  
}
\hfill $\diamond$
\end{example}

Proposition~\ref{Prop-gapG} does not allow to show the $G$-operators associated to Examples~\ref{ex-Odd} and \ref{ex-Even} are gapped because $A$ and $H$ are not necessarily unitary, even though all other operators of the fuzzy tori are unitary which, moreover, commute with each other. Of course, the situation of Example~\ref{ex-Even} is dealt with in detail in the proof of Theorem~\ref{theo-Intro} given in Section~\ref{sec-EvenPL}. The next result generalizes Proposition~\ref{Prop-gapG} to invertible operators $A_1,\ldots,A_d\in\BM(\Hh)$ that form a fuzzy $d$-torus. This also provides the gap estimate of Theorem~\ref{theo-Intro}, albeit with considerably worse constants.

\begin{proposi}
\label{Prop-gapGA}
If $A_1, \ldots, A_d\in \BM(\Hh)$ form a fuzzy $d$-torus of width $\delta\in(0,\frac{1}{2}]$ the selfadjoint operator $G=G_{\{1,\ldots,d\}}(A_1,\ldots,A_d)$ satisfies 
\begin{align}
G^2
\;\geq\;
\big(1\,-\,24\,d^2\delta\big)\one\;.
\label{eq-gapGA}
\end{align}
In particular, $G$ is invertible if $\delta<\frac{1}{24\,d^2}$.
\end{proposi}

\noindent {\bf Proof.}
Using the Clifford relations one has
\begin{align}
\nonumber
G^2
&
\;=\;
\left(\sum_{j=1}^d \Im m(A_j)^2+\left((d-1)\one-\sum_{j=1}^d \Re e(A_j)\right)^2\right)
+\,\sum_{1\leq j<i\leq d}[\Im m(A_j),\Im m(A_i)]\otimes \gamma_j\gamma_i
\\
\nonumber
& \;\;\;\;\;\;\;\;\;+\,\sum_{j=1}^d[\Im m(A_j),(d-1)\one-\sum_{i=1}^d \Re e(A_i)]\otimes \gamma_j\gamma_{d+1}\\
\nonumber
&\geq\;
\left(\sum_{j=1}^d \Im m(A_j)^2+\left((d-1)\one-\sum_{j=1}^d \Re e(A_j)\right)^2\right)\\
\label{eq-boundGA1}
&
\;\;\;\;\;\;\;\;\;-\,\left(\sum_{1\leq j<i\leq d}\|[\Im m(A_j),\Im m(A_i)]\|\,-\,\sum_{j,i=1}^d\|[\Im m(A_j),\Re e(A_i)]\|\right)
\;.
\end{align}
%
Using that $\spec(|A_j|)\subset [(1-\delta)^\frac{1}{2}, (1+\delta)^\frac{1}{2}]$ and $\delta\leq\frac{1}{2}$, one checks that 
\begin{equation}
\label{eq-NormBound}
\|A_j\|\,=\,\||A_j|\|\,\leq\,(1+\delta)^{\frac{1}{2}}\,\leq\,1+\frac{\delta}{2}\;,
\qquad
\|A_j^{-1}\|^2\,\leq \,(1-\delta)^{-1}\,\leq\,1+2\delta
\;.
\end{equation}
Therefore
\begin{align}
\nonumber
\|[A_i,A_j^*]\|
&\;=\;
\|A_j^{-1}(A_jA_iA_j^*A_j-A_jA_j^*A_iA_j)A_j^{-1}\|\\
\nonumber
&\;\leq\;
\|A_j^{-1}\|^2\,\|A_jA_i(A_j^*A_j-\one)+[A_j,A_i]+(\one-A_jA_j^*)A_iA_j\|\\
\label{eq-kommAA*}
&\;\leq\;
(1+2\delta) \big(2(1+\delta)\delta+\delta\big)
\;\leq\;8\,\delta
\;.
\end{align}
This leads to 
\begin{align*}
\|[\Im m (A_j), \Im m (A_i)]\|
& \leq
\frac{1}{4}\left(\big\|[A_j,A_i]\big\|+\big\|[A_i,A_j^*]\big\|+\big\|[A_i^*,A_j]\big\|+\big\|[A_j^*,A_i^*]\big\|\right)
\leq\frac{1}{2}(\delta+8\delta)\leq 5\,\delta\;.
\end{align*}
In the same way one shows the bound
\begin{align*}
\|[\Im m (A_j), \Re e (A_i)]\|
\;\leq\;
\frac{1}{4}\left(\big\|[A_j,A_i]\big\|+\big\|[A_i,A_j^*]\big\|+\big\|[A_j,A_i^*]\big\|+\big\|[A_i^*,A_j^*]\big\|\right)
\;\leq\;
5\,\delta
\;,
\end{align*}
for $j\neq i$. For $j=i$ a slightly better estimate holds
$$
\|[\Im m (A_j), \Re e (A_j)]\|
\;=\;
\frac{1}{2}\|A_jA_j^*-A_j^*A_j\|\;\leq\;\delta\;.
$$
Inserting this into \eqref{eq-boundGA1} leads to 
\begin{align}
\label{eq-boundGA2}
G^2
\,-\,
\left(\sum_{j=1}^d \Im m(A_j)^2+\left((d-1)\one-\sum_{j=1}^d \Re e(A_j)\right)^2\right)\,\geq\,
-\left(d\delta+15\binom{d}{2}\delta\right)\one
\,\geq\,
-9\,d^2\,\delta\,\one\;.
\end{align}
It thus remains to prove a lower bound on the term in the parenthesis. Using
$$
\Im m(A_j)^2+\Re e(A_j)^2
\;=\;
\frac{1}{2}(A_jA_j^*+A_j^*A_j)
\;\geq\;
\one-\frac{1}{2}\|(A_jA_j^*-\one)+(A_j^*A_j-\one)\|\one
\;\geq\;(1-\delta)\one
\,,
$$
one finds
\begin{align}
& \sum_{j=1}^d \Im m(A_j)^2+\left((d-1)\one-\sum_{j=1}^d \Re e(A_j)\right)^2
\nonumber
\\
&\;\geq\;
d(1-\delta)\one+(d-1)^2\one-2(d-1)\sum_{j=1}^d \Re e(A_j)\,+\!\sum_{1\leq j<i\leq d}^d \!\big(\Re e(A_j)\Re e(A_i)+\Re e(A_i)\Re e(A_j)\big)
\nonumber
\\
\label{eq-boundSumReA}
&\;=\;
\one+\sum_{1\leq j<i\leq d}\big((\one-\Re e(A_j))(\one-\Re e(A_i))+(\one-\Re e(A_i))(\one-\Re e(A_j))\big)-d\delta\one\;.
\end{align}
One directly checks that 
$$
\frac{1}{2}(A_j-\one)(A_j-\one)^*-(\one-\Re e(A_j))\;=\;\frac{1}{2}(A_jA_j^*-\one)
$$
and therefore, by the first part of \eqref{eq-deffuzzytorus2},
$$
\big\|\frac{1}{2}(A_j-\one)(A_j-\one)^*-(\one-\Re e(A_j))\big\|\;\leq\;\frac{1}{2}\,\delta\;.
$$
Moreover, using \eqref{eq-NormBound} 
$$
\|\one-\Re e(A_j)\|\;\leq\;1+(1+\delta)^\frac{1}{2}\,\leq\,\frac{5}{2}\;,
\qquad
\|A_j-\one\|^2\;\leq\;\big(1+(1+\delta)^\frac{1}{2}\big)^2\,\leq\,5\;.
$$
Using the last two bounds one gets
\begin{align*}
&
\big\|\frac{1}{4}(A_j-\one)(A_j-\one)^*(A_i-\one)(A_i-\one)^*-(\one-\Re e (A_j))(\one-\Re e (A_i))\big\|
\\
&\;\leq\;
\big\|\frac{1}{2}(A_j-\one)(A_j-\one)^*\big(\frac{1}{2}(A_i-\one)(A_i-\one)^*-(\one-\Re e(A_i))\big)\big\|
 \\ 
 &
\;\;\;\;\;\;\;\;
+\big\|\big(\frac{1}{2}(A_j-\one)(A_j-\one)^*-(\one-\Re e(A_j))\big)(\one-\Re e(A_i)\big\|
\\
& 
\;\leq\;
\frac{1}{2}\,5\,\frac{1}{2}\,\delta\,+\,\frac{1}{2}\,\delta\,\frac{5}{2}
\;=\;
\frac{5}{2}\,\delta
\;.
\end{align*}
Inserting this into \eqref{eq-boundSumReA} leads to 
\begin{align*}
&
\sum_{j=1}^d \Im m(A_j)^2+\left((d-1)\one-\sum_{j=1}^d \Re e(A_j)\right)^2
\\
&
\;\geq\;
\one+\frac{1}{4}\!\sum_{1\leq j<i\leq d}\!\!\big((A_j-\one)(A_j-\one)^*(A_i-\one)(A_i-\one)^*
+(A_i-\one)(A_i-\one)^*(A_j-\one)(A_j-\one)^*\big)
\\
&
\hspace{2cm}
-d\,\delta\,\one-2\,\binom{d}{2}\,\frac{5}{2}\,\delta
\\
&
\;\geq\;
\one+\frac{1}{4}\sum_{1\leq j<i\leq d}\Big( (A_j-\one)(A_i-\one)(A_i-\one)^*(A_j-\one)^*
\\
&
\hspace{2cm}
+(A_j-\one)[(A_j-\one)^*,(A_i-\one)](A_i-\one)^*+(A_j-\one)(A_i-\one)[(A_j-\one)^*,(A_i-\one)^*]
\\
&
\hspace{2cm}
+(A_i-\one)(A_j-\one)(A_j-\one)^*(A_i-\one)^*+(A_i-\one)[(A_i-\one)^*,(A_j-\one)](A_j-\one)^*
\\
&
\hspace{2cm}
+(A_i-\one)(A_j-\one)[(A_i-\one)^*,(A_j-\one)^*]\Big)
-\frac{7}{2}\,d^2\,\delta\,\one
\;.
\end{align*}
The first and fourth summand are non-negative and can thus be left out for a lower bound. In the other four summands, the commutators reduce to $[A_j^*,A_i]$ or $[A_j^*,A_i^*]$ which can be bounded directly by \eqref{eq-deffuzzytorus2}  or by \eqref{eq-kommAA*}. Using, moreover, again $\|A_j-\one\|^2\leq 5$, one thus obtains
\begin{align*}
\sum_{j=1}^d \Im m(A_j)^2+\left((d-1)\one-\sum_{j=1}^d \Re e(A_j)\right)^2
&
\;\geq\;
\Big(1-\frac{5}{4}\,\binom{d}{2}\,(8\delta+\delta+8\delta+\delta)\Big)\one\,-\frac{7}{2}\,d^2\,\delta\,\one
\\
&
\;\geq\;\Big(1-15\,d^2\, \delta\Big)\one
\;.
\end{align*}
Combining with \eqref{eq-boundGA2} one obtains \eqref{eq-gapGA}.
\hfill $\Box$

\vspace{.2cm}

The next result shows that a graded fuzzy $d$-torus can always be reduced to a suitably associated ungraded fuzzy $d$-torus.

\begin{proposi}
\label{Prop-Greduce}
Let $A_1,\ldots,A_d,A_{d+1}=A_{d+1}^*\in\Aa^\sim$ be graded fuzzy $d$-torus of a sufficiently small width $\delta\leq\frac{1}{2}$. Let $P$ denote the Riesz projection on the positive spectrum of $A_{d+1}$. Then $PA_1P,\ldots,PA_dP$ form a fuzzy $d$-torus of width $6\delta$ on the Hilbert space $P\Hh$. Setting $G^P_I=G_{I}(PA_1P,\ldots,PA_dP)$ for $I\subset\{1,\ldots,d\}$, the operator $\widehat{G}_I=\widehat{G}_{I}(A_1,\ldots,A_d,A_{d+1})$ is homotopic to $G^P_I\oplus (\one-P)\gamma_{d+1}$ inside the invertible operators, still for $\delta$ sufficiently small. In particular,
$$
\Sig(G^P_I)
\;=\;
\Sig(\widehat{G}_I)
\;.
$$
\end{proposi}

\noindent {\bf Proof.}
As $\|A_{d+1}^2-\one\|\leq \delta$ by assumption, $A_{d+1}$ is close to a symmetry and its spectrum is separated into two intervals $[-\sqrt{1+\delta},-\sqrt{1-\delta}]$ and $[\sqrt{1-\delta},\sqrt{1+\delta}]$, see {\it e.g.} \eqref{eq-NormBound}. Let $P$ denote the Riesz projection associated to $[\sqrt{1-\delta},\sqrt{1+\delta}]$. Then $\|A_{d+1}-(2P-\one)\|\leq 1-\sqrt{1-\delta}\leq\delta$ by the spectral mapping theorem. Moreover, the other conditions in \eqref{eq-deffuzzytorus2} imply 
$$
\|[A_j,P]\|
\;\leq\;
\max_{z\in\partial B_1(\pm 1)}\|(z\one-A_{d+1})^{-1}\|^2\|[A_j,A_{d+1}]\|\;\leq 2\,\delta
\;,
$$
so that 
\begin{equation}
\label{eq-Pquasidiag}
\big\|A_j-PA_jP-(\one-P)A_j(\one-P)\big\|
\;=\;
\big\|(\one-P)A_jP+PA_j(\one- P)\big\|
\;\leq\;2\,\delta
\;.
\end{equation}
Moreover, since $\|A_j\|=\|A_j^*A_j\|^{\frac{1}{2}}\leq (1+\delta)^{\frac{1}{2}}\leq 1+\frac{1}{2}\delta$,
$$
\big\|[PA_jP,PA_iP]\big\|
\;\leq\;
\big\|P[A_j,A_i]P\big\|
+
\big\|P[A_j,P]A_iP\big\|
+
\big\|[P[A_i,P]A_jP\big\|
\;\leq \; 6\,\delta
\;,
$$
the operators $PA_1P,\ldots,PA_dP$ forms an (ungraded) fuzzy $d$-torus on the Hilbert space $P\Hh$ of width $6\delta$. (Similarly, also  $(\one-P)A_1(\one-P),\ldots,(\one-P)A_d(\one-P)$ is an (ungraded) fuzzy $d$-torus on the Hilbert space $(\one-P)\Hh$ of width $6\delta$, but this torus will not be used.) The associated $G$-operator is denoted by $G^P_I$, see the statement of the proposition. One then has, due to \eqref{eq-Pquasidiag},
$$
\big\|
\widehat{G}_I\,-\,
G^{P}_I\oplus \widetilde{G}^{\one-P}_I\big\|
\;\leq\;5\,d\,\delta
\;,
$$
where 
$$
\widetilde{G}^{\one-P}_I
\!=
\sum_{j\in I}(\one-P) \Im m(A_{j})(\one-P)\otimes \gamma_j+\Big((|I|+1)(\one-P)-\sum_{j\in I}(\one-P)\Re e(A_{j})(\one-P)\Big)\otimes \gamma_{|I|+1}
\;.
$$
Let us stress that this is {\it not} the $G$-operator $G^{\one-P}_I$ on $(\one-P)\Hh$ associated to the fuzzy $d$-torus $(\one-P)A_1(\one-P),\ldots,(\one-P)A_d(\one-P)$ for $I\subset\{1,\ldots,d\}$ by the definition \eqref{eq-GDef}, simply because one of the summands is $(|I|+1)(\one-P)\otimes \gamma_{|I|+1}$ rather than $(|I|-1)(\one-P)\otimes \gamma_{|I|+1}$. It hence remains to show that $\widetilde{G}^{\one-P}_I$ is homotopic to $(\one-P)\otimes \gamma_{|I|+1}$ inside the invertible operators. This follows directly from the next lemma, applied to the fuzzy $|I|$-torus associated to $I$.
\hfill $\Box$

\begin{lemma}
\label{lem-TrivialCase}
Given a fuzzy $d$-torus $A_1,\ldots,A_d$ of a sufficiently small width $\delta$, the operator
$$
\widetilde{G}
\,=\,
\sum_{j=1}^{d}\Im m(A_{j})\otimes \gamma_j+\Big((d+1)\one-\sum_{j=1}^{d}\Re e(A_{j})\Big)\otimes \gamma_{d+1}
$$
is homotopic to $\gamma_{d+1}$ inside the invertible operators.
\end{lemma}

\noindent {\bf Proof.} The homotopy will be given by the straight-line path $t\in[0,1]\mapsto \widetilde{G}(t)=(1-t)\gamma_{d+1}+t\widetilde{G}$. Explicitly
$$
\widetilde{G}(t)
\,=\,
\gamma_{d+1}
+t
\Big(\sum_{j=1}^{d}\Im m(A_{j})\otimes \gamma_j+\Big(\sum_{j=1}^{d}(\one-\Re e(A_{j}))\Big)\otimes \gamma_{d+1}\Big)
\;.
$$
Now all commutators $[\Im m(A_{j}), \Re e(A_{i})]$ and $[\Im m(A_{j}), \Im m(A_{i})]$ are of order $\Oo(\delta)$, see the proof of Proposition~\ref{Prop-gapGA}. Hence
$$
\widetilde{G}(t)^2
\,=\,
\one
+2t\sum_{j=1}^{d}(\one-\Re e(A_{j}))
+
t^2 
\Big(\sum_{j=1}^{d}\Im m(A_{j})^2+\big(\sum_{j=1}^{d}(\one-\Re e(A_{j}))\big)^2\Big)
\,+\,\Oo(\delta)
\;.
$$
But up to errors of order $\Oo(\delta)$, one also has $\one-\Re e(A_{j})\geq 0$. Hence $\widetilde{G}(t)^2\geq \one+\Oo(\delta)$, which implies the claim. Note that the claim merely reflects that the maps $g_{d,d+1}$ defined in Appendix~\ref{app-MapsExplicit} have a vanishing mapping degree and are hence homotopic to the identity.  
\hfill $\Box$

\vspace{.2cm}

Now that the crucial property that the operators $G_I$ are gapped is proved for fuzzy tori of sufficiently small width, it is possible to extract topological information from them.  Recall that elements of the $K$-group $K_0(\Aa)$ are homotopy equivalence classes of either projections or equivalently invertible selfadjoints in matrix algebras of $\Aa$, and that $K_1(\Aa)$ are homotopy equivalence classes of invertibles in matrix algebras of $\Aa$ or equivalent equivalence classes of selfadjoint invertible which anti-commute with some symmetry in the matrix degrees of freedom (see \eqref{eq-OddReduce}, or {\it e.g.} \cite{LS1} for some further explanations of this). Based on Proposition~\ref{Prop-gapGA} and Lemma~\ref{lem-GGrading} one therefore has the following. 

\begin{coro}
\label{coro-Reduce}
Let $A_1,\ldots,A_d$ be a fuzzy $d$-torus in a $C^*$-algebra $\Aa$ of sufficiently small width. For any index set $I\subset\{1,\ldots,d\}$, one then obtains a class $[G_I]_0\in K_0(\Aa)$ if $I$ is of even cardinality and a class  $[G_I]_1\in K_1(\Aa)$ if $I$ is of odd cardinality. For a graded fuzzy torus one obtains $[\widehat{G}_I]_0\in K_0(\Aa)$ for $I$ of even cardinality, and $[\widehat{G}_I]_1\in K_1(\Aa)$ for $I$ of odd cardinality. In the latter case, the classes can also be represented by the $G$-operators $G^P_I$ of the $P$-restricted fuzzy tori given in {\rm Proposition~\ref{Prop-Greduce}}.
\end{coro}

If the algebra $\Aa$ is given by matrices (or compact operators) and $A_1,\ldots,A_d$ is a fuzzy $d$-torus of matrices, then one can read out its $K$-theoretic content using the half-signatures $\frac{1}{2}\,\Sig(G_I)\in\ZM$ with $I\subset\{1,\ldots,d\}$ being of even cardinality. Note that the number of such $I$ is given by $\sum_{j=0}^{\lfloor \frac{d}{2} \rfloor}\binom{d}{2j}=2^{d-1}$. For a graded fuzzy $d$-torus of matrices, the invariants are given by $\frac{1}{2}\,\Sig(\widehat{G}_I)=\frac{1}{2}\,\Sig(G^P_I)$, again for $I\subset\{1,\ldots,d\}$ of even cardinality so that there are again $2^{d-1}$ invariants. 

\vspace{.2cm}

\noindent {\bf Proof} of Theorem~\ref{theo-Intro2}. One only has to apply Corollary~\ref{coro-Reduce} to Example~\ref{ex-Even}.
\hfill $\Box$

\vspace{.2cm}

In the case of two almost commuting unitaries $U_1,U_2$ (see Example~\ref{eq-NCtorus2}) satisfying that $\Sig(G_{\{1,2\}}(U_1,U_2))=0$, it is known \cite{ELP} that they can be deformed into two commuting unitaries. Hence one can expect the integers $\frac{1}{2}\,\Sig(G_{\{1,2\}}(U_1,U_2))$ to fully classify all fuzzy $2$-tori. We even suspect that all fuzzy matrix tori are completely classified by the signature invariants constructed above, provided $G_\emptyset$ is suitably defined:

\vspace{.2cm}

\noindent {\bf Conjecture:} {\it Two fuzzy $d$-tori of matrices having the same signature invariants $\Sig({G}_I)$ for all $I\subset\{1,\ldots,d\}$ of even cardinality can be homotopically deformed into each other without closing the gaps of any of the $G_I$. Similarly, graded fuzzy $d$-tori of matrices are classified by the invariants $\frac{1}{2}\,\Sig(\widehat{G}_I)=\frac{1}{2}\,\Sig(G^P_I)$, again for $I\subset\{1,\ldots,d\}$ of even cardinality.}

\appendix

\section{Mapping degree versus Chern number}
\label{app-Degree}

For the convenience of the reader, this appendix provides a detailed proof of the connection between mapping degree of a differentiable function $f: \TM^d\to\SM^d$ on an even-dimensional torus and the Chern number of an associated matrix-valued projection $P_f:\TM^d\to\CM^{d'\times d'}$ with $d'=2^{\frac{d}{2}}$.  This fact is used in Appendix~\ref{app-MapsExplicit} which is crucial for understanding the motivation for the periodic spectral localizer and the $G$-operators in Section~\ref{sec-Fuzzy}. While the main statement, Corollary~\ref{coro-ChernDeg} below, is certainly well-known in the community, we could not localize a detailed proof.

\vspace{.2cm}

Let $d$ be even and let us denote the restrictions of the euclidean coordinate functions to $\SM^d$ by $x_j: \SM^d\to\RM$ for $j\in\{1,\ldots, d+1\}$. More precisely,
$$
x_j(y)
\;=\;
y_j\;, \qquad 
y\,=\,
\begin{pmatrix}
y_1\\
\vdots\\
y_{d+1}
\end{pmatrix}\in\SM^d\;.
$$
Furthermore, let $f_j=x_j\circ f: \TM^d\to\RM$ be $j$th component of the function $f$ for $j\in\{1,\ldots, d+1\}$. Let $\gamma_1,\ldots,\gamma_{d+1}\in\CM^{d'\times d'}$ be an irreducible self-adjoint representation of the Clifford algebra with the convention that
\begin{equation}
\label{eq-convClifford}
\gamma_1\cdots\gamma_{d+1}
\;=\;
\imath^{\frac{d}{2}}\one\;.
\end{equation}
Then let us define the map $P_f: \TM^d\to \CM^{d'\times d'}$ by 
%
$$
P_f(k)
\;=\;
\frac{1}{2}\Big(\sum_{j=1}^{d+1}f_j(k)\gamma_j+\one\Big)
\;.
$$
Then $P_f(k)$ is an (orthogonal) projection for all $k\in \TM^d$. Its exterior derivative is $dP_f= \frac{1}{2}\sum_{j=1}^{d+1}\gamma_jdf_j$, a matrix-valued $1$-form on $\TM^d$. Then let us set
$$
\omega_f
\;=\; 
\Tr\big(P_f(dP_f\wedge dP_f)^{\wedge\frac{d}{2}}\big)
\;,
$$
which is a $d$-form on $\TM^d$. Using \eqref{eq-convClifford} as $\Tr(\one)=2^{\frac{d}{2}}$, it is explicitly given by
\begin{align*}
\omega_f
&\;=\; 
C_d\sum_{j=1}^{d+1}(-1)^{j+1}f_jdf_1\wedge\ldots\wedge df_{j-1}\wedge df_{j+1}\wedge\ldots \wedge df_{d+1}
\;,
\qquad
C_d
\,=\,
\frac{1}{2}\frac{1}{2^d}d!2^{\frac{d}{2}}\imath^{\frac{d}{2}}
\end{align*}
Similarly, let us define another projection-valued map $P^W: \SM^d\to \CM^{d'\times d'}$ (called the Weyl projection) by
$$
P^W(p)
\;=\; 
\frac{1}{2}\Big(\sum_{j=1}^{d+1}x_j(p)\gamma_j+\one\Big)
\;.
$$
Similar as above, there is an associated $d$-form on $\SM^d$ given by
\begin{align*}
\omega^W
\,=\,
\Tr\big(P^W(dP^W\wedge dP^W)^{\wedge\frac{d}{2}}\big)
\,=\,
C_d
\sum_{j=1}^{d+1}(-1)^{j+1}x_jdx_1\wedge\ldots\wedge dx_{j-1}\wedge dx_{j+1}\wedge\ldots \wedge dx_{d+1}\;.
\end{align*}

\begin{proposi} 
For any differentiable map $f: \TM^d \to \SM^d$, the differential form $\omega_f$ on $\TM^d$ equals the pullback of the differential form $\omega^W$ by $f$:
$$
\omega_f
\;=\; 
f^*\omega^W\;.
$$
\end{proposi}

\noindent {\bf Proof.}
For a point $k\in \TM^d$ let $\eta_1, \ldots, \eta_d: (-a,a)\to \TM^d$ for $a>0$ represent tangent vectors of $\TM^d$ at the point $k$, namely $\eta_j$ is a differentiable curve fulfilling $\eta_j(0)=k$ for all $j\in\{1, \ldots, d\}$. The tangent vector represented by $\eta_j$ is denoted by $[\eta_j]$. Then one has to show
\begin{equation}
\label{eq-omegaf=PBomegaw}
\omega_f([\eta_1],\ldots,[\eta_d])
\;=\; 
\omega^W(df_k([\eta_1]), \ldots, df_k([\eta_d]))\;,
\end{equation}
where $df_k$ denotes the differential of $f$ at the point $k$. A direct computation shows
\begin{align*}
&\omega_f([\eta_1],\ldots,[\eta_d])\\
&\;=\; 
C_d
\sum_{j=1}^{d+1}(-1)^{j+1}f_jdf_1\wedge\ldots\wedge df_{j-1}\wedge df_{j+1}\wedge\ldots \wedge df_{d+1}([\eta_1],\ldots,[\eta_d])\\
&\;=\;
C_d
\sum_{j=1}^{d+1}(-1)^{j+1}f_j(k)\frac{1}{d!}\sum_{\sigma\in S_d}df_1([\eta_{\sigma(1)}])\cdots df_{j-1}([\eta_{\sigma(j-1)}]) df_{j+1}([\eta_{\sigma(j)}])\cdots df_{d+1}([\eta_{\sigma(d)}])\\
&\;=\;
\frac{C_d}{d!}
\sum_{j=1}^{d+1}(-1)^{j+1}f_j(k)\sum_{\sigma\in S_d}(f_1\circ\eta_{\sigma(1)})'(0)\cdots (f_{j-1}\circ\eta_{\sigma(j-1)})'(0)
\\
&
\hspace{6cm}
(f_{j+1}\circ\eta_{\sigma(j)})'(0)\cdots (f_{d+1}\circ\eta_{\sigma(d)})'(0)
\;.
\end{align*}
In a similar manner one checks
\begin{align*}
&\omega^W(df_k([\eta_1]), \ldots, df_k([\eta_d]))\\
&
\;=\; 
C_d
\sum_{j=1}^{d+1}(-1)^{j+1}x_jdx_1\wedge\ldots\wedge dx_{j-1}\wedge dx_{j+1}\wedge\ldots \wedge dx_{d+1}(df_k([\eta_1]), \ldots, df_k([\eta_d]))\\
&
\;=\;
C_d\sum_{j=1}^{d+1}(-1)^{j+1}x_j(f(k))\frac{1}{d!}\sum_{\sigma\in S_d}dx_1(df_k[\eta_{\sigma(1)}])\cdots dx_{j-1}(df_k[\eta_{\sigma(j-1)}]) 
\\
&
\hspace{6cm}
dx_{j+1}(df_k[\eta_{\sigma(j)}])\cdots dx_{d+1}(df_k[\eta_{\sigma(d)}])\\
&\;=\;
\frac{C_d}{d!}
\sum_{j=1}^{d+1}(-1)^{j+1}f_j(k)\sum_{\sigma\in S_d}(x_1\circ f\circ \eta_{\sigma(1)})'(0)\cdots (x_{j-1}\circ f\circ \eta_{\sigma(j-1)})'(0) \\
&
\hspace{6cm}
(x_{j+1}\circ f\circ \eta_{\sigma(j)})'(0)\cdots (x_{d+1}\circ f\circ \eta_{\sigma(d)})'(0)\\
&\;=\;
\frac{C_d}{d!}
\sum_{j=1}^{d+1}(-1)^{j+1}f_j(k)\sum_{\sigma\in S_d}(f_1\circ\eta_{\sigma(1)})'(0)\cdots (f_{j-1}\circ\eta_{\sigma(j-1)})'(0)\\
&
\hspace{6cm}
(f_{j+1}\circ\eta_{\sigma(j)})'(0)\cdots (f_{d+1}\circ\eta_{\sigma(d)})'(0)\;.
\end{align*}
Therefore \eqref{eq-omegaf=PBomegaw} holds and the claim follows.
\hfill $\Box$

\vspace{.2cm}

Now by a well-known pullback formula ({\it e.g.} tom Dieck's lecture notes \cite{TD} contain a detailed proof) one has
$$
\int_{\TM^d}\omega_f\;=\;\deg(f)\int_{\SM^d}\omega^W\;,
$$
where $\deg(f)$ is the mapping degree of $f$, generically defined as the sum over all preimages  (of a fixed point) of the signs of the determinants of the Jacobians. Next recall ({\it e.g.} \cite{PSB}) the definition of the $d$th Chern number of a differentiable projection-valued map $P:\Mm^d\to \CM^{L\times L}$ on a $d$-dimensional manifold $\Mm$:
$$
\Ch_d(P)
\;=\;(-1)^{\frac{d}{2}}(\tfrac{1}{2\pi \imath })^{\frac{d}{2}}\frac{1}{\frac{d}{2}!}\int_{\Mm^d}\Tr\big(P(dP)^d\big)
\;.
$$
Then for $\omega_f=\Tr\big(P_f(dP_f)^d\big)$ as above, one obtains
$$
\Ch_d(P_f)
\;=\;\deg(f)(-1)^{\frac{d}{2}}(\tfrac{1}{2\pi i })^{\frac{d}{2}}\frac{1}{\frac{d}{2}!}\int_{\SM^d}\omega^W
\;=\; \deg(f)\, \Ch_d(P^W)\;.
$$
It hence remains to compute $\Ch_d(P^W)$ which is again well-known:
$$
\Ch_d(P^W)
\;=\;(-1)^{\frac{d}{2}}\;.
$$
({\it E.g.} \cite{CSB,SSt} contains a detailed computation.) Summing up, one concludes:

\begin{coro}
\label{coro-ChernDeg}
For $d$ even and a smooth map $f:\TM^d\to\SM^d$, the $d$th Chern number of  $P_f$ is 
$$
\Ch_d(P_f)
\;=\;(-1)^{\frac{d}{2}}\deg(f)\;.
$$
\end{coro}

\section{Mapping degree of some maps from torus to sphere}
\label{app-MapsExplicit}

This appendix is about the mapping degrees of the maps $g_{d,m}: \TM^d \to \RM^{d+1}$ with $d$ even and $m\in\RM$ given by
\begin{equation}
\label{eq-defmapg}
g_{d,m}\big(e^{\imath\theta_1}, \ldots, e^{\imath\theta_d}\big)
\;=\; 
\Big(
\sin(\theta_1),
\ldots,
\sin(\theta_d),
m-\sum_{n=1}^d\cos(\theta_n)\Big)
\;,
\end{equation}
where $\theta_n\in[0,2\pi)$ for $n\in\{1, \ldots, d\}$ and $(e^{\imath\theta_1}, \ldots, e^{\imath\theta_d})\in\TM^d$.  It can readily be checked that  the vector on the r.h.s. does not vanish if and only if  $m\in\RM \setminus\{-d, -d+2, \ldots, d-2,d\}$. For such $m$, let us then set
\begin{equation}
\label{eq-defmapf}
f_{d,m}\big(e^{\imath\theta_1}, \ldots, e^{\imath\theta_d}\big)
\;=\;
\big\|g_{d,m}\big(e^{\imath\theta_1}, \ldots, e^{\imath\theta_d}\big)\big\|^{-1}
g_{d,m}\big(e^{\imath\theta_1}, \ldots, e^{\imath\theta_d}\big)
\;.
\end{equation}
Then $f_{d,m}: \TM^d \to \SM^d$ is a map onto the unit sphere with same mapping degree $\deg(f_{d,m})=\deg(g_{d,m})$. By Corollary~\ref{coro-ChernDeg}, this mapping degree is equal to the $d$-th Chern number $\Ch_d(P_{f_{d,m}})$. These Chern numbers were computed in Section~2.2.4 of \cite{PSB} by analyzing the changes of the Chern numbers at the transition points $\{-d, -d+2, \ldots, d-2,d\}$. The argument involves rather delicate singular integrals, and this was revisited in detail in \cite{SSt}. Here a direct alternative argument based on the computation of the mapping degree of $g_{d,m}$ is provided. Let us stress that the map $g_{d,d-1}$ is at the root of the construction of the periodic spectral localizer and the $G$-operators associated to fuzzy tori. It  leads to a Chern number $\Ch_d(P_{f_{d,d-1}})=1$.

\vspace{.2cm}

To compactify notations, let us set $e^{\imath{\bm \theta}}=(e^{\imath\theta_1}, \ldots, e^{\imath\theta_d})\in\TM^d$ and ${\bm x}=(x_1,\ldots,x_{d+1})\in\RM^{d+1}$ as well as $\hat{\bm x}=(x_1,\ldots,x_{d})\in\RM^{d}$. Furthermore, let $\TM^d$ be equipped with the orientation inherited from the atlas $(U_{\bf j},\varphi_{\bf j})_{{\bf j}\in {\bf J}}$ with ${\bf J}=\{0,1\}^d$ given by, for $ {\bf j}=(j_1,\ldots,j_d)\in {\bf J}$,
\begin{align*}
U_{\bf j}
& \,=\,
\big\{e^{\imath{\bm \theta}}\in\TM^d : \theta_n\in[0,2\pi)\setminus\{\tfrac{\pi}{2}+j_n\pi\} \;\text{for}\; n\in\{1, \ldots, d\}\big\}
\;,
\end{align*}
and the charts
\begin{align*}
\varphi_{\bf j}\big(e^{\imath{\bm \theta}}\big)
\,=\,
\big((-1)^{|{\bf j}|}\tfrac{\cos(\theta_1)}{1+(-1)^{j_1}\sin(\theta_1)},\tfrac{\cos(\theta_2)}{1+(-1)^{j_2}\sin(\theta_2)}, \ldots,\tfrac{\cos(\theta_d)}{1+(-1)^{j_d}\sin(\theta_d)}\big) 
\;,
\end{align*}
where $|{\bf j}|=\sum_{n=1}^dj_n$. Note that the factor $(-1)^{|{\bf j}|}$ in the first component assures that all charts are positively oriented.
Moreover, let $\SM^d$ be equipped with orientation inherited from the atlas $(V_j, \psi_j)_{j\in\{1,2\}}$ given by
\begin{align*}
V_1
\,=\,
\left\{
{\bm x}\in\SM^d\subset \RM^{d+1} : x_{d+1}\neq1\right\}\;,
\quad
V_2
\,=\,
\left\{
{\bm x}
\in\SM^d\subset \RM^{d+1} : x_{d+1}\neq-1\right\}\;,
\end{align*}
and
\begin{align*}
\psi_1\big({\bm x}\big)
\;=\;
\big(\tfrac{x_1}{1-x_{d+1}}, \ldots,\tfrac{x_d}{1-x_{d+1}}\big) \;,
\qquad
\psi_2\big({\bm x}\big)
\;=\;
\big(\tfrac{x_2}{1+x_{d+1}}, \tfrac{x_1}{1+x_{d+1}}, \tfrac{x_3}{1+x_{d+1}}, \ldots,\tfrac{x_d}{1+x_{d+1}}\big)\;.
\end{align*}

\begin{proposi} 
If $\TM^d$ is equipped with the orientation inherited from the atlas $(U_{\bf j},\varphi_{\bf j})_{{\bf j}\in{\bf J}}$ and $\SM^d$ is equipped with the orientation inherited from the atlas $(V_j, \psi_j)_{j\in\{1,2\}}$ the degree of the map $f_{d,m}$ defined by \eqref{eq-defmapf} is given by
$$
\deg(f_{d,m})
\;=\;
\begin{cases}
\sum_{k=0}^\frac{d-n-1}{2}(-1)^k\binom{d}{k}\,, & n\in(0,d+1)\cap2\NM+1, m\in(n-1,n+1)\;,\\
\sum_{k=0}^\frac{d-n-1}{2}(-1)^{k+1}\binom{d}{k}\,, & n\in(0,d+1)\cap2\NM+1, m\in(-n-1,-n+1)\;,\\
0\,, & m\in (-\infty,-d)\cup(d,\infty)\,.
\end{cases}
$$
\end{proposi}

\noindent {\bf Proof.} By the homotopy invariance of the mapping degree, it is sufficient to consider the case $m\in \{-d-1, -d+1, \ldots, d+1\}$. For the case $m=d+1$, one can consider the maps
$$
g_{d,d+1,t}\big(e^{\imath{\bm \theta}}\big)
\;=\; 
\Big(
t \sin(\theta_1),
\ldots,
t \sin(\theta_d),
1-t\sum_{n=1}^d(\cos(\theta_n)-1)\Big)
\;,
$$
where $t\in[0,1]$. Then $g_{d,d+1,1}=g_{d,d+1}$ and $g_{d,d+1,0}(e^{\imath{\bm \theta}})=p_N$ where $p_N=(0,\ldots,0,1)\in\SM^d$ is the north pole. Now the norm satisfies
$$
\|g_{d,d+1,t}\big(e^{\imath{\bm \theta}}\big)\|^2
\;=\;
t^2\sum_{j=1}^d\sin^2(\theta_j)
+
1+ 2t\sum_{j=1}^d(1-\cos(\theta_j))
+t^2\Big(\sum_{j=1}^d(1-\cos(\theta_j))\Big)^2
\;\geq\;1
\;.
$$
This implies that $t\in [0,1]\mapsto f_{d,d+1,t}=\|g_{d,d+1,t}\|^{-1}g_{d,d+1,t}$ is a homotopy from $f_{d,d+1}$ to a constant map. Hence the mapping degree vanishes. (Note that this argument is essentially reproduced for fuzzy tori in Lemma~\ref{lem-TrivialCase}.) Let us next focus on the case $m\in\{1, 3, \ldots, d-1\}$. The other points are dealt with in a similar manner (with the north instead of the south pole used in the argument below). Consider the south pole given by $p_S=(0,\ldots,0,-1)\in\SM^d$. Its inverse image is
\begin{align*}
&f_{d,m}^{-1}(p_S)
\;=\;
\big\{e^{\imath{\bm \theta}}\in\TM^d : \theta_n \in \{0,\pi\} \;\text{for}\; n\in\{1, \ldots, d\}, \theta_j=\pi \;\text{for at most}\; \tfrac{d-m-1}{2} \;\text{many}\; j \big\}\;.
\end{align*}
Then, for ${\bf 1}=(1,\ldots,1)\in{\bf J}$,
$$
\psi_1\circ f_{d,m} \circ \varphi_{\bf 1}^{-1}: \RM^d\setminus\big\{\hat{\bm x}\in\RM^d: x_n\in\{\pm1\},\#\{j\in\{1,\ldots,d\}: x_j=-1\}> \tfrac{d-m-1}{2}\big\} \;\to \;\RM^d
$$
is given by
\begin{align*}
(\psi_1\circ f_{d,m} \circ \varphi_{\bf 1}^{-1})\big(\hat{\bm x}\big)
\;=\;
\left(\tfrac{1-x_1^2}{x_1^2+1}\big(1-m+2\sum_{n=1}^d\tfrac{x_n}{x_n^2+1}\big)^{-1},\ldots, \tfrac{1-x_d^2}{x_d^2+1}\big(1-m+2\sum_{n=1}^d\tfrac{x_n}{x_n^2+1}\big)^{-1}\right).
\end{align*}
Its $k$th component is
\begin{equation}
\label{eq-psifvarphi-1at_-p}
(\psi_1\circ f_{d,m} \circ \varphi_{\bf 1}^{-1})\big(\hat{\bm x}\big)_k
\;=\;
\frac{(1-x_k^2)\prod_{n\neq k}(x_n^2+1)}{(1-m)\prod_{n=1}^d(x_n^2+1)+2\sum_{n=1}^dx_n\prod_{j\neq n}(x_j^2+1)}\;.
\end{equation}
Let us denote the Jacobian matrix of this map by 
$$
J=
J_{\psi_1\circ f_{d,m} \circ \varphi_{\bf 1}^{-1}}:  \RM^d\setminus\big\{\hat{\bm x}\in\RM^d: x_n\in\{\pm1\},\#\{j\in\{1,\ldots,d\}: x_j=-1\}> \tfrac{d-m-1}{2}\big\}\; \to\; \RM^d\;.
$$
To determine the mapping degree of $f$ (at the point $p_S$) it is sufficient to compute the restriction of $J$ to $\MM=\varphi_{\bf 1}(f^{-1}(p_S))$ explicitly given by
$$
\MM
\;=\;
\big\{\hat{\bm x}\in\RM^d: x_n\in \{-1,1\} \;\text{for} \; n\in\{1, \ldots, d\},  x_j=-1 \;\text{for at most}\; \tfrac{d-m-1}{2} \;\text{many}\; j\big\}\;.
$$
One directly checks that the off-diagonal entries of $J|_\MM$ vanish, namely $J(\hat{\bm x})_{k,l}=0$ for all $\hat{\bm x}\in \MM$ and $k,l\in \{1, \ldots, d\}$, $k\neq l$. The diagonal entries of $J|_\MM$ are
\begin{equation}
\label{eq-diagentryJ}
J(\hat{\bm x})_{k,k}\;=\;\frac{-x_k2^d}{((1-m)2^d+2^d\sum_{n=1}^d x_j)^2}
\end{equation}
for $\hat{\bm x}\in \MM$ and $k\in \{1, \ldots, d\}$. Therefore $\sgn (J(\hat{\bm x})_{k,k})=-\sgn(x_k)$ and 
$$
\sgn(\det(J(\hat{\bm x})))
\;=\;
 (-1)^{\#\{j\in\{1, \ldots, d\}: x_j=1\}}
 \;=\;
 (-1)^{\#\{j\in\{1, \ldots, d\}: x_j=-1\}}
\;,
$$
where the last equality holds because $d$ is even. Because 
$$
\#\big\{\hat{\bm x}\in\RM^d: x_n\in \{-1,1\} \;\text{for} \; n\in\{1, \ldots, d\},  x_j=-1 \;\text{for}\; k \;\text{many}\; j\big\}
\;=\;
\binom{d}{k}
\;,
$$
one has 
$$
\deg(f_{d,m}, p_S)
\;=\;
\sum_{k=0}^\frac{d-n-1}{2}(-1)^k\binom{d}{k}
\;.
$$
As the mapping degree is independent of the point at which the preimage is taken (provided it has a finite preimage), this shows the claim for $m\in\{1, 3, \ldots, d-1\}$. 
\hfill $\Box$

\vspace{.2cm}

\noindent {\bf Acknowledgements:}  T.L.\ acknowledges support from the National Science Foundation, grant DMS-2110398. The work of H.S.-B. was supported by the DFG (Deutsche Forschungsgesellschaft) under grant SCHU 1358/8-1. Data sharing is not applicable to this article as no datasets were generated or analyzed during the current study. The authors have no competing interests to declare that are relevant to the content of this article.


\end{document}